\documentclass[pra, aps,nopacs,twocolumn,floatfix,superscriptaddress,amsmath,   longbibliography]{revtex4-1}
\usepackage{graphicx}
\usepackage{graphics}
\usepackage{braket}
\usepackage{mathtools}
\usepackage[titletoc]{appendix}
\usepackage{appendix}
\usepackage{float}
\usepackage{bm}
\usepackage{amssymb}

\usepackage[dvipsnames]{xcolor}
\definecolor{darkblue}{rgb}{0.0, 0.0, 0.75}

\usepackage{color}
\usepackage[colorlinks=true,
            linkcolor=darkblue,
            urlcolor=darkblue,
            citecolor=darkblue]{hyperref}

	\usepackage[normalem]{ulem} %makes underlining possible: \uline \uwave \sout
	\definecolor{mgreen}{RGB}{1,123,0}

\def \bk{{\bf k}}
\def \br{{\bf r}}

\def \bp{{\bf p}}
\def \bq{{\bf q}}
\def \md{\mathrm{d}}

\def \mum{\mu \mathrm{m}}

\def \m2D{\mathrm{2D}}
\def \mB{\mathrm{B}}
\def \mms{\mathrm{ms}}
\def \mmms{\mathrm{mm/s}}

\def \mkin{\mathrm{kin}}
\def \mpot{\mathrm{pot}}
\def \mint{\mathrm{int}}

\def \mCOM{\text{COM}}

\def \mnK{\text{nK}}

\def \mcyl{\text{cyl}}

\def \mrelax{\text{relax}}

\def \meff{\text{eff}}

\def \tk{\tilde{k}}
\def \td{\tilde{d}}

\def \tv{\tilde{v}}

\def \mHz{\mathrm{Hz}}

\def \ms{\mathrm{s}}

\def \cH{\mathcal{H}}

\def \bk{\mathbf{k} }
\def \br{\mathbf{r} }

\newcommand*\diff{\mathop{}\!\mathrm{d}}

\begin{document}
\title{Superfluidity of a laser-stirred Bose-Einstein condensate}
\author{Hannes Kiehn}
\affiliation{Zentrum f\"ur Optische Quantentechnologien and Institut f\"ur Laserphysik, Universit\"at Hamburg, 22761 Hamburg, Germany}
\author{Vijay Pal Singh}
\affiliation{Zentrum f\"ur Optische Quantentechnologien and Institut f\"ur Laserphysik, Universit\"at Hamburg, 22761 Hamburg, Germany}
\affiliation{Institut f\"ur Theoretische Physik, Leibniz Universit\"at Hannover, Appelstra{\ss}e 2, 30167 Hannover, Germany}
\author{Ludwig Mathey}
\affiliation{Zentrum f\"ur Optische Quantentechnologien and Institut f\"ur Laserphysik, Universit\"at Hamburg, 22761 Hamburg, Germany}
\affiliation{The Hamburg Centre for Ultrafast Imaging, Luruper Chaussee 149, Hamburg 22761, Germany}
\date{\today}
%
%%%------------------------------------------ ABSTRACT ----------------------------------------------------------------------------------
%
\begin{abstract}
We study superfluidity of a cigar-shaped Bose-Einstein condensate (BEC) by stirring it with a Gaussian potential oscillating back and forth along the axial dimension of the condensate, motivated by experiments of C. Raman \textit{et al.} Phys. Rev. Lett. \textbf{83}, 2502 (1999).  
Using classical-field simulations and perturbation theory we examine the induced heating rate, based on the total energy of the system,  as a function of the stirring velocity $v$. 
We identify the onset of dissipation by a sharply increasing heating rate above a velocity $v_c$, which we define as the critical velocity. 
We show that $v_c$ is influenced by the oscillating motion, the strength of the stirrer, the temperature and the inhomogeneous density of the cloud. 
This results in a vanishing $v_c$ for the parameters similar to the experiments, which is inconsistent with the measurement of nonzero $v_c$. 
However, if the heating rate is based on the thermal fraction after a $100\, \mms$ equilibration time, our simulation recovers the experimental observations. 
We demonstrate that this discrepancy is due to the slow relaxation of the stirred cloud and dipole mode excitation of the cloud.
\end{abstract}

\maketitle
%
%---------------------------------------------------------------Introduction: BEGIN -------------------------------------------------------
%
\section{Introduction}

An intriguing phenomenon of quantum liquids is the emergence of superfluidity, which was first observed in liquid helium \cite{Kapitza1938, Misener1938}.  
This superfluid behavior can be tested by dragging an object through a stationary fluid and examining the dissipationless flow around this object. 
For an object moving at a velocity $v$, dissipationless flow occurs below a certain critical velocity $v_c$. 
Above this velocity, dissipation occurs via the creation of excitations such as phonons and vortices. 
An estimate of the critical velocity is given by the Landau criterion $v_c = \mathrm{min} \bigl( E_\bk/(\hbar \bk) \bigr)$, 
where $E_\bk$ is the energy of an excitation with momentum $\hbar \bk$ \cite{Pethick2008}.
This predicts the critical velocity to be the roton velocity for liquid helium 4 and the pair-breaking velocity for liquid helium 3, consistent with the measurements in Refs. \cite{Allum1977, Pickett1986, McClintock1995}.

  With the advent of Bose-Einstein condensates (BECs) of dilute gases, the study of superfluidity was expanded to a wide range of quantum liquids.  
A pioneering technique of local perturbation, namely, laser stirring was employed for probing superfluidity in these systems. 
This includes the measurements of superfluidity in cigar-shaped condensates \cite{Ketterle1999, Ketterle2001, Ketterle2000, Engels2007}, two-dimensional (2D) condensates \cite{Dalibard2012}, and 3D condensates \cite{Weimer2015}.
Given this broad variety of systems, different origins of the onset of dissipation have been identified, such as vortex-antivortex creation  \cite{Singh2017} for the experiment in Ref. \cite{Dalibard2012}, and phonon creation \cite{Singh2016} for the experiment in Ref. \cite{Weimer2015}. 
Vortex shedding and quantized vortices were observed in Refs.  \cite{Dalibard2000, Inouye2001, KetterleVortex, KetterleVortex2, Shin2015, Shin2016, Shin2018}.
Other stirring experiments include Refs. \cite{ Ketterle2007, Salomon2014, Pfau2018, Sobirey2021, Shin2021}. 
Laser stirring is also employed to generate controlled vortex distributions \cite{Gauthier2019, Wilson2021, Kwon2021}.
A number of theoretical studies was reported in Refs. \cite{Adams1999, Zwerger2000, Adams2000, Walsworth2000, Amy2014, Davis1, Davis2, Singh2021SF, Zheng2019, Chevy2019, Ronning2020}.

    In  Ref. \cite{Ketterle1999}  a cigar-shaped cloud of  $^{23}$Na atoms was stirred with a blue-detuned laser beam oscillating back and forth along the axial direction. 
Using a strong beam, the superfluid response was measured based on the thermal fraction determined as a function of the stirring velocity, which is varied by the stirring displacement at a fixed stirring frequency. 
The resulting critical velocity was $v_c/c_\mB=0.25$. 
In follow-up experiments with improved calorimetry \cite{Ketterle2001}, the critical velocity was determined using low stirring velocities and for a colder system, yielding  $v_c/c_\mB=0.07 \pm 0.01$.   
 
\begin{figure}[]
\includegraphics[width=1.0\linewidth]{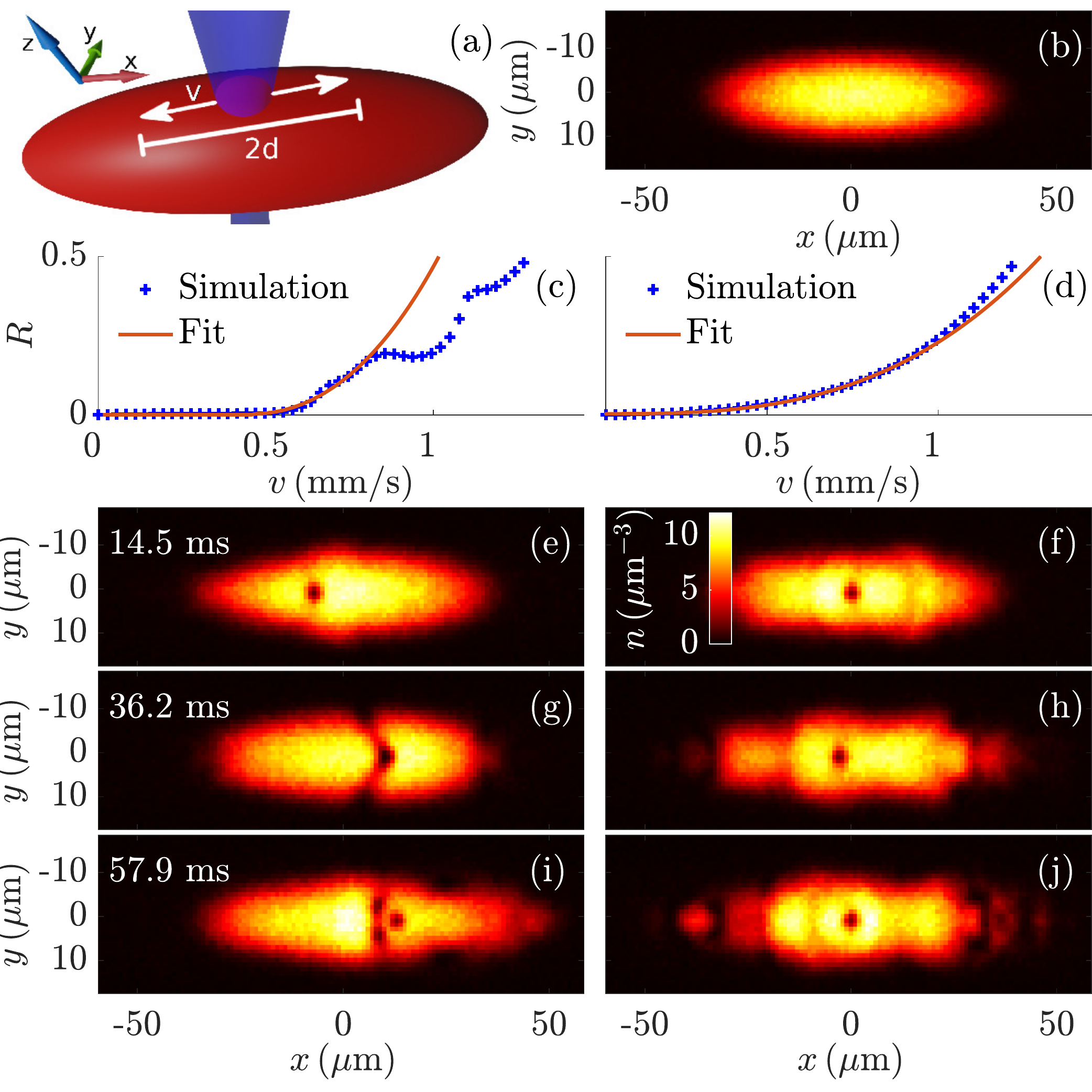}
\caption{(a) Sketch of the stirring process: A blue-detuned laser beam (blue) stirs a cigar-shaped cloud (red). 
The stirring velocity $v=4fd$ is determined by the oscillation frequency $f$  and the displacement $d$ along the $x$ direction.
(b) Simulated density cloud with the density at the trap center $n_0= 10.7\, \mum^{-3}$ and the temperature $T/T_c=0.75$.  
The dimensionless heating rate $R$ determined as a function of $v$, for fixed (c) $d=20\, \mum$ and  
(d) $f=56\, \mHz$.  The beam strength is $V_0/\mu=2$, where $\mu$ is the mean-field energy at the trap center.
(e, g, i) Dynamics of the central plane density for $v=0.8\, \mmms$ and stirring times  of $21.7$, $43.4$, and $65.2\, \mms$ corresponding to panel (c), while the results in panels (f, h, j) correspond to panel (d).
The critical velocity $v_c$, determined via fitting (continuous lines), is $0.46 \pm 0.01\, \mmms$ and $0$ for the heating rate in (c) and (d), respectively. }
\label{Fig1}
\end{figure}

  In this paper we study the experiments of Ref. \cite{Ketterle1999} from a theoretical perspective. 
The stirring process is described in Fig. \ref{Fig1}(a), 
where the stirring velocity is determined by $v= 4fd$, with $f$ being the frequency and $d$ the displacement. 
The condensate dynamics are simulated with a classical-field method, using parameters that are comparable to the experiments. 
The equilibrium cloud is shown in Fig. \ref{Fig1}(b).
Using simulation and analytical results for a homogeneous system we provide insight into the reduction of the critical velocity due to the oscillating motion, the temperature, and the strong stirring potential. 
We derive the heating rate perturbatively in the stirring term, 
which shows excellent agreement with the simulations and describes the oscillating-motion broadening analytically. 
Furthermore, we confirm the measurements of nonzero critical velocity \cite{Ketterle1999} if the superfluid response is 
based on the thermal fraction of the non-equilibrated cloud.  
We expand on this by analyzing the slow relaxation of the stirred cloud, which occurs on a timescale of $5\, \ms$ much longer than the used $0.1 \, \ms$ relaxation time in the experiments. 
Finally, we discuss the dynamical regimes of phonon- and vortex-induced dissipation and show that the latter occurs for strong stirring potentials and large displacements of the stirrer.

In Figs. \ref{Fig1}(c) - (j) we present an example for the stirring process and the analysis of its heating rate.
Given that the velocity is determined via $v = 4fd$, its magnitude can be tuned either by varying $f$ and keeping $d$ fixed, or by varying $d$ and keeping $f$ fixed. The heating rate of the first scenario is shown in Fig. \ref{Fig1}(c), and snapshots of the density distribution are shown in Figs. \ref{Fig1}(e, g, i). The heating rate of the second scenario, and snapshots of the density distribution are shown in Figs. \ref{Fig1}(d, f, h, j). We note that the second scenario does not give a nonzero critical velocity, determined via fitting, whereas the first scenario does. Furthermore, we note the strong excitation of the atom cloud that is visible in the snapshots, which only displays a slow relaxation to a thermalized state, as we show below. 
These two examples exemplify key ingredients of our subsequent discussion.

 This paper is organized as follows.
In Sec. \ref{sec:sim} we describe our simulation method. 
In Sec. \ref{sec:hom} we derive the analytical heating rate and show its comparison to the simulations. 
In Sec. \ref{sec:V0st} we point out the mechanism of vortex-induced dissipation for strong stirrers. 
In Sec. \ref{sec:temp} we discuss the reduction of the onset of dissipation due to temperature. 
In Sec. \ref{sec:trap} we show the simulations for trapped clouds. 
In Sec. \ref{sec:relax} we describe the slow relaxation of the stirred trapped cloud 
and we conclude in Sec. \ref{sec:con}

\begin{figure*}[]
\includegraphics[width=1.0\linewidth]{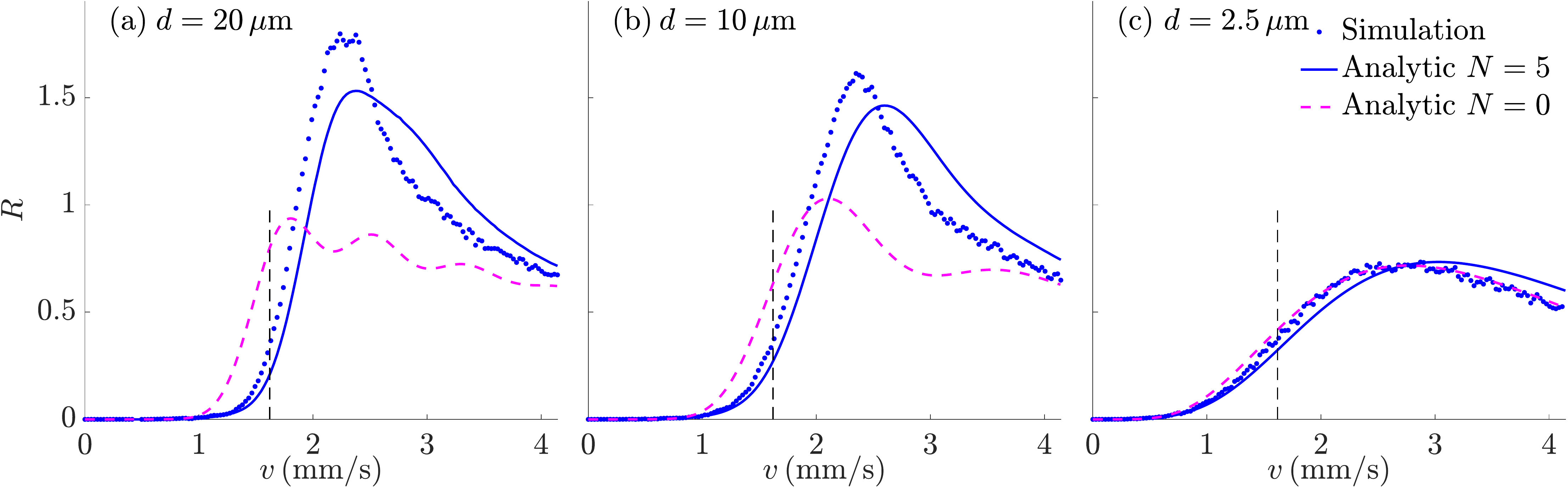} 
\caption{Simulated heating rate $R$ (dots) for a homogeneous condensate at $T/T_c=0.05$, which is induced by stirring the condensate with a weak stirrer oscillating back and forth along the $x$ direction. 
Panels (a-c) show $R$ as a function of the stirring velocity $v=4fd$, for $d=20$, $10$ and $2.5\, \mum$, respectively. 
The continuous lines are the Bogoliubov estimates of Eq. \ref{eq:Rate}. 
The dashed lines correspond to Eq. \ref{eq:RateCos2}, where the motion is approximated by a cosine stirring motion. 
The vertical dashed lines denote the Bogoliubov phonon velocity $v_\mB=1.62\, \mmms$.  }
\label{Fig2}
\end{figure*}
\section{Simulation method}\label{sec:sim}
 Motivated by the experiments of Ref. \cite{Ketterle1999}, we study superfluidity of a cigar-shaped cloud of $^{23}$Na atoms confined in a 3D harmonic trap. 
We simulate the stirring dynamics using the c-field method of Ref. \cite{Singh2016}. 
The unperturbed system is described by the Hamiltonian  
\begin{multline}\label{eq:Hamil}
    \hat{H}_0 = \int \mathrm{d}\br\Bigl[ \frac{\hbar^2}{2m} \nabla \hat{\psi}^\dagger({\bf r})  \cdot \nabla \hat{\psi}({\bf r}) +V(\br)\hat{\psi}^\dagger (\br) \hat{\psi}(\br)\\ + \frac{g}{2} \hat{\psi}^\dagger({\bf r})\hat{\psi}^\dagger({\bf r})\hat{\psi}({\bf r})\hat{\psi}({\bf r}) \Bigr].
\end{multline}
$\hat{\psi}$ ($\hat{\psi}^\dagger$) is the bosonic annihilation (creation) operator.
The interaction constant $g=4\pi a_s \hbar^2/m$ is determined by the s-wave scattering length $a_s$ and the atom mass $m$. 
The external potential represents a harmonic trap $V(\br)=m \bigl( \omega_x^2 x^2+\omega_\perp^2 (y^2+z^2) \bigr)/2$, 
where $\omega_x $ and $\omega_\perp$ are the trap frequencies in the axial and the radial direction, respectively.   
To describe laser stirring we add a time-dependent term 
\begin{equation}
  \hat{H}_s(t)=\int \mathrm{d} \br V(\br,t)\hat{n}(\br).
\end{equation}
$\hat{n}(\br)$ is the density operator at the location $\br=(x, y, z)$. 
$V(\br, t)$ is the Gaussian potential 
\begin{equation}\label{eqn:Stirring potential}
V(\br,t)=V_0 \exp\left(-\frac{[(x-x(t))^2+(y-y_0)^2]}{2\sigma^2}\right),
\end{equation}
which is constant along the $z$ direction. $V_0$ is the strength and $\sigma$ is the width. 
This potential is centered around $x(t)$ and $y_0$. 
We set $y_0$ to be at the cloud center and vary $x(t)$ periodically  at a constant velocity $v$ between the points  $x=-d$ and $x=d$ as described in Fig. \ref{Fig1}(a).
A constant velocity means that the velocity flips sign at the turning points instantaneously. 
$d$ is chosen such that the distance between the turning points is $2 d$. 
The potential travels a distance $4d$ during one oscillation period that is given by $T=4 d/v$.  
The frequency of the oscillation is $f=v/(4 d)$. 
The periodic motion $x(t) = x(t+T)$ is described as
\begin{equation}\label{eq:xt}
  x(t) = \begin{cases}
      -d+v t &   0\leq t\leq T/2  \\
    3d-v t &   T/2 < t < T,
 \end{cases}
\end{equation}
which  is characterized by two parameters $v$ and $d$. 

    For the numerical simulations we discretize space on a lattice of size $N_x \times N_y \times N_z$ and a discretization length $l= 1 \, \mum$. 
We choose $l$ such that it is smaller than the healing length $\xi=\hbar/\sqrt{2 m g n}$ and the de Broglie wavelength $\lambda= \sqrt{2\pi \hbar^2/(m k_\mB T)}$ \cite{Castin2003}. $n$ is the density, $T$ the temperature, and $k_\mB$ the Boltzmann constant. 
We replace the operators $\hat{\psi}$ in Eq. \ref{eq:Hamil} and in the equations of motion by complex numbers $\psi$. 
We sample the initial states in a grand canonical ensemble of temperature $T$ and chemical potential $\mu$ via a classical Metropolis algorithm. 
For the system parameters we choose  $(\omega_x, \, \omega_\perp) = 2 \pi \times (9, \, 32.5)\, \mathrm{Hz}$, the total number of atoms $N \approx1.48\times10^5$, and $T/T_c=0.75$. 
The critical temperature is estimated by $T_c \approx 0.94 \hbar \bar{\omega}N^{1/3}/k_\mB$, 
where $\bar{\omega} = (\omega_x \omega_\perp^2)^{1/3}$ is the geometric mean of the trap frequencies \cite{Pethick2008}. 
This results in the trap central density $n_0=10.7\, \mum^{-3}$, and Thomas-Fermi diameters of $84.8$ and $23.5\, \mum$ in the axial and radial directions, respectively. 
Since the number of atoms is one order of magnitude smaller than the experiments, our system represents a scaled down version of the experiment.  
For comparison, we also consider homogeneous clouds of the average density $n=10\,  \mum^{-3}$ and various $T/T_c$. 
Here, $T_c$ is estimated by $T_c \approx 3.31 \hbar^2 n^{2/3}/(m k_\mB)$ \cite{Pethick2008}.
We employ a lattice of $150 \times 50 \times 50$ ($128\times 128 \times 8$) sites for simulations of the trapped (homogeneous) cloud.

    Specifically, we implement the stirring protocol as follows. 
We choose a set of stirring parameters $V_0$, $\sigma$, $d$, and $f$. 
For homogeneous (trap) simulations we use  $2 \sigma=2\, \mum$ ($2\, \mum$ and $3.25\, \mum$) and $V_0$ in the range $V_0/\mu=0.08 -5$ ($1 -6.36$). 
The mean-field energy $\mu= g n$ is determined using the average (maximum) density of homogeneous (trapped) system.
The stirring protocol is the following. 
We turn on $V_0$ linearly over $4 \, \mms$, stir the cloud for $72 \, \mms$, and then turn off $V_0$ linearly again. 
We repeat this for a desired set of stirring parameters and the ensemble. 
We calculate an ensemble average of the energy $E=\braket{H_0}$ using Eq. \ref{eq:Hamil}, with the operators  $\hat{\psi}$ replaced by complex numbers $\psi$. 
During stirring, $E$ increases linearly with time and from its slope we determine the heating rate $\diff{E}/\diff{t}$.
The dimensionless heating rate is obtained by $R=\hbar(\diff{E}/\diff{t})/(N_\mcyl V_0^2)$, where  $N_\mcyl=\pi N \sigma^2/A$ is the number of atoms in a cylinder of length $L_z$ and radius $\sigma$. $A$ is the system area. 
We show $R$ for various system and stirring parameters below.

\section{Homogeneous BEC at low temperature}\label{sec:hom}

  To understand and isolate various features of the stirring process we first consider a homogeneous condensate at a low temperature of $T/T_c=0.05$. 
We stir this system with a weak stirrer following the protocol described in Sec. \ref{sec:sim}. 
We use $V_0/\mu= 0.087$ and $2\sigma=2.0\, \mum$.
We determine the heating rate $R$  for various values of $v=4fd$ by varying $f$ and keeping $d$ constant.
In Fig. \ref{Fig2} we show $R(v)$ for fixed $d= 20$, $10$, and $2.5\, \mum$.
We find two regimes in the heating rate. 
For $v$ much lower than the Bogoliubov estimate of the phonon velocity $v_\mB=\sqrt{g n/m} = 1.62\, \mmms$, the stirrer produces almost no heating of the system, showcasing the superfluid behaviour of the condensate.  
For high $v$ close to $v_\mB$,  excitations are created in the system, leading to a sharp increase of the heating rate.
To characterize this transition between the two regimes, we determine the onset of dissipation by a critical velocity $v_c$ using the fitting function
\begin{equation}\label{eqn:fitfunction}
\left(\frac{\md E}{\md t}\right)_{\text{Fit}}=A \frac{(v^2-v_c^2)^2}{v}+B,
\end{equation}
with $A$, $B$, and $v_c$ being the free parameters \cite{Astrakharchik2004}. 
For $R(v)$ in Fig. \ref{Fig2}, this fitting function yields $v_c = 1.26$, $1.03$, and $0.37\, \mmms$ for $d= 20$, $10$, and $2.5\, \mum$, respectively. 
The magnitude of $v_c$ is below $v_\mB$ and decreases with decreasing $d$. We explain this below analytically.

  We derive an analytical expression for the heating rate by considering the stirring term as a perturbative term \cite{Singh2016}.  
We use Bogoliubov theory at zero temperature and solve the equations of motion. The total Hamiltonian of the system is $\hat{H}(t)=\hat{H}_0+\hat{H}_s(t) $. For the unperturbed uniform system, the Bogoliubov approximation gives the diagonalized Hamiltonian
\begin{align}\label{eq:HBog}
 \hat{\cH}_0 = \sum_{\bm{k}} \hbar \omega_k \hat{b}_{\bm{k}}^\dagger \hat{b}_{\bm{k}}.
\end{align}
$\hat{b}_{\bm{k}}$ ($\hat{b}_{\bm{k}}^\dagger$) is the Bogoliubov annihilation (creation) operator and $\hbar \omega_k = \sqrt{\epsilon_k (\epsilon_k + 2 m v_\mB^2)}$ is the Bogoliubov dispersion. $\epsilon_k =\hbar^2 k^2/(2 m) $ is the free-particle dispersion. 
In momentum space, the stirring term becomes
\begin{align}
 \hat{H}_s(t) = \sum_{\bk} V_{\bk}(t) \hat{n}_{\bk}.
\end{align}
$V_{\bk}(t)$ and $\hat{n}_{\bk}$ are the Fourier transforms of the stirring potential and the density operator, respectively. 
To calculate $V_{\bk}(t)$ we expand the periodic motion in Eq. \ref{eq:xt} into a Fourier series of the form 
$x(t) = \sum_{\lambda=0}^{N} a_\lambda \cos(\omega_\lambda t)$, with $a_\lambda$ being the amplitudes and $\omega_\lambda $ the  frequencies. 
Within the Bogoliubov approximation we use $\hat{n}_\bk \approx \sqrt{N_0} (u_k+v_k)(\hat{b}_{-\bk}^\dagger+\hat{b}_{\bk})$, where $N_0$ is the number of condensate atoms, and $u_k$ and $v_k$ are the Bogoliubov parameters.
This results in 
\begin{align} \label{eq:Hst}
 \hat{\cH}_s(t) = \sum_{\bk} V_{\bk}(t)  \sqrt{N_0} (u_k+v_k)(\hat{b}_{-\bk}^\dagger+\hat{b}_{\bk}), 
\end{align}
with 
\begin{align}
 V_{\bk}(t) = V_{\bk} \prod_{\lambda=0}^{N} \exp \bigl[ i k_x a_\lambda \cos(\omega_\lambda t) \bigr].
\end{align}
$V_{\bk} = (2\pi V_0 \sigma^2/A) \times  \delta_{k_z} \exp(-k^2 \sigma^2/2) \exp(i k_y y_0)$ is the time-independent term. 
We solve the dynamical evolution created by Eqs. \ref{eq:HBog} and \ref{eq:Hst} using the ansatz
\begin{align} \label{eq:ansatz}
 \hat{b}_\bk(t)=e^{-i \omega_k t} \hat{b}_\bk+A_\bk(t),
\end{align}
where $A_\bk(t)$ is given by (see Appendix \ref{sec:appRate})
\begin{align}
 A_{\bk}(t) &= - \frac{2 i}{\hbar}  V_{\bk} \sqrt{N_0} (u_k + v_k) \smashoperator[l]{\sum_{\nu_0,\dots,\nu_N=-\infty}^\infty} \prod_{\lambda=0}^N \left[i^{\nu_\lambda}J_{\nu_\lambda}(k_x a_\lambda)\right]  \nonumber \\ 
& \quad \times e^{-i(\omega_k-\omega_{\meff}) t/2} \frac{\sin[(\omega_k-\omega_{\meff}) t/2]}{\omega_k-\omega_{\meff}}.
\end{align}
$\omega_{\meff}$ is defined as $\omega_{\meff} =  \sum_{\lambda=0}^N \nu_\lambda \omega_\lambda$ and $J_\nu(x)$ is the Bessel function of the first kind of order $\nu$.
With Eq. \ref{eq:ansatz} we determine the expectation value of the energy $\braket{E(t)}=\sum_{\bk} \hbar \omega_k \braket{\hat{b}_{\bk}^\dagger(t)\hat{b}_{\bk}(t)}$. This results in the energy change  $\braket{\Delta E(t)}=\sum_{\bk} \hbar \omega_k |A_\bk(t)|^2$. 
The time derivative of the energy change yields the heating rate (see Appendix \ref{sec:appRate})
\begin{equation}\label{eq:Rate}
\frac{\diff{E}}{\diff{t}} =  \frac{2 \pi}{ \hbar} \sum_{\bk}  \omega_k (u_k + v_k)^2 N_0|V_{\bk}|^2 \smashoperator[lr]{\sum_{\nu_0,\nu_1,\nu'_1,\dots,\nu_N,\nu'_N=-\infty}^\infty} B(k_x) \delta(\omega_k-\omega_{\meff}), 
\end{equation}
where $B(k_x)$ is given in Appendix \ref{sec:appRate} and includes a product of various Bessel functions of the first kind of order $\nu$. Eq. \ref{eq:Rate} cannot be reduced analytically and thus we solve it numerically following the method described in Appendix \ref{sec:appRate}. 
To gain insight into the result in Eq. \ref{eq:Rate}, we take the first-order term in the Fourier expansion $x(t) = a \cos(\omega t)$, which is a cosine motion instead of a linear-oscillating motion originally considered in Eq. \ref{eq:xt}.
This results in the heating rate (Appendix \ref{sec:appRate})
\begin{align}\label{eq:rate_cos}
\frac{\diff{E}}{\diff{t}} &=  \frac{2 \pi}{ \hbar} \sum_{\bk, \nu}  \omega_k (u_k + v_k)^2 N_0|V_{\bk}|^2  J_\nu^2(ak_x) \delta(\omega_k-\nu \omega ).
\end{align}
This expression is similar to the heating rate for a circular stirring motion in Ref. \cite{Singh2016}. 
$a= 8 d/\pi^2$ represents the 1D equivalent of a radius. 
To solve Eq. \ref{eq:rate_cos} we replace the $\bk$ sum by an integral and obtain
\begin{equation} \label{eq:rate_cos1}
R = 2 \sum_{\nu=- \infty}^\infty \tk_c^2 e^{-\tk_c^2} \frac{\sqrt{\tk_c^4 + 4 \tv_\mB^2 \tk_c^2 } }{\tk_c^2+ 2 \tv_\mB^2} \int_0^1 d\kappa \frac{ J_\nu^2 (8 \kappa \tk_c \td/\pi^2)} { \sqrt{1- \kappa^2}},
\end{equation}
with 
\begin{align}
\tk_c = \bigl( -2 \tv_\mB^2 + \sqrt{4 \tv_\mB^4 + \pi^2 \nu^2 \tv^2 \tv_\mB^2/\tilde{d}^2 }  \bigr)^{1/2}.
\end{align}
$R=\hbar(\diff{E}/\diff{t})/(N_\mcyl V_0^2)$ is the dimensionless heating rate. $\tv=v/v_\mB$, $\tv_\mB= v_\mB m \sigma/\hbar$, $\td=d/\sigma$, and $\tk=k \sigma$ are dimensionless parameters.  Eq. \ref{eq:rate_cos1} can be solved, giving  
\begin{align}\label{eq:RateCos2}
R &= 2 \sum_{\nu=1}^\infty  \tk_c^2 e^{-\tk_c^2} \frac{\sqrt{\tk_c^4 + 4 \tv_\mB^2 \tk_c^2 } }{\tk_c^2+ 2 \tv_\mB^2} (8 \tk_c \tilde{d}/\pi^2)^{2 \nu} \Gamma^2\Bigl( \nu+ \frac{1}{2} \Bigr) \nonumber \\
& \times F\Bigl(  \nu+ \frac{1}{2},  \nu+ \frac{1}{2}; \nu+1, \nu+1, 2\nu+1;  -(8 \tk_c \tilde{d}/\pi^2)^2  \Bigr).
\end{align}
$F(\ell, \ell; m, m, n  ; x)$ is the regularized generalized hypergeometric function. We calculate the estimates of Eqs. \ref{eq:Rate} and \ref{eq:RateCos2} as a function of $v= 4fd$ for the same stirring parameters as in the simulations and show these results in Fig. \ref{Fig2}. 
The results of the estimates and the simulation are consistent for $d=2.5\, \mum$ and deviate for large $d$. 
The estimate of Eq. \ref{eq:Rate} shows a maximum at a higher $v$ than in the simulation. 
The low-$v$ regime of the heating rate is well captured by Eq. \ref{eq:Rate} for all $d$, providing an accurate estimate of the onset of dissipation, 
whereas the estimate of Eq. \ref{eq:RateCos2} captures this low-$v$ regime only for small $d$.

\begin{figure}[]
\includegraphics[width=1\linewidth]{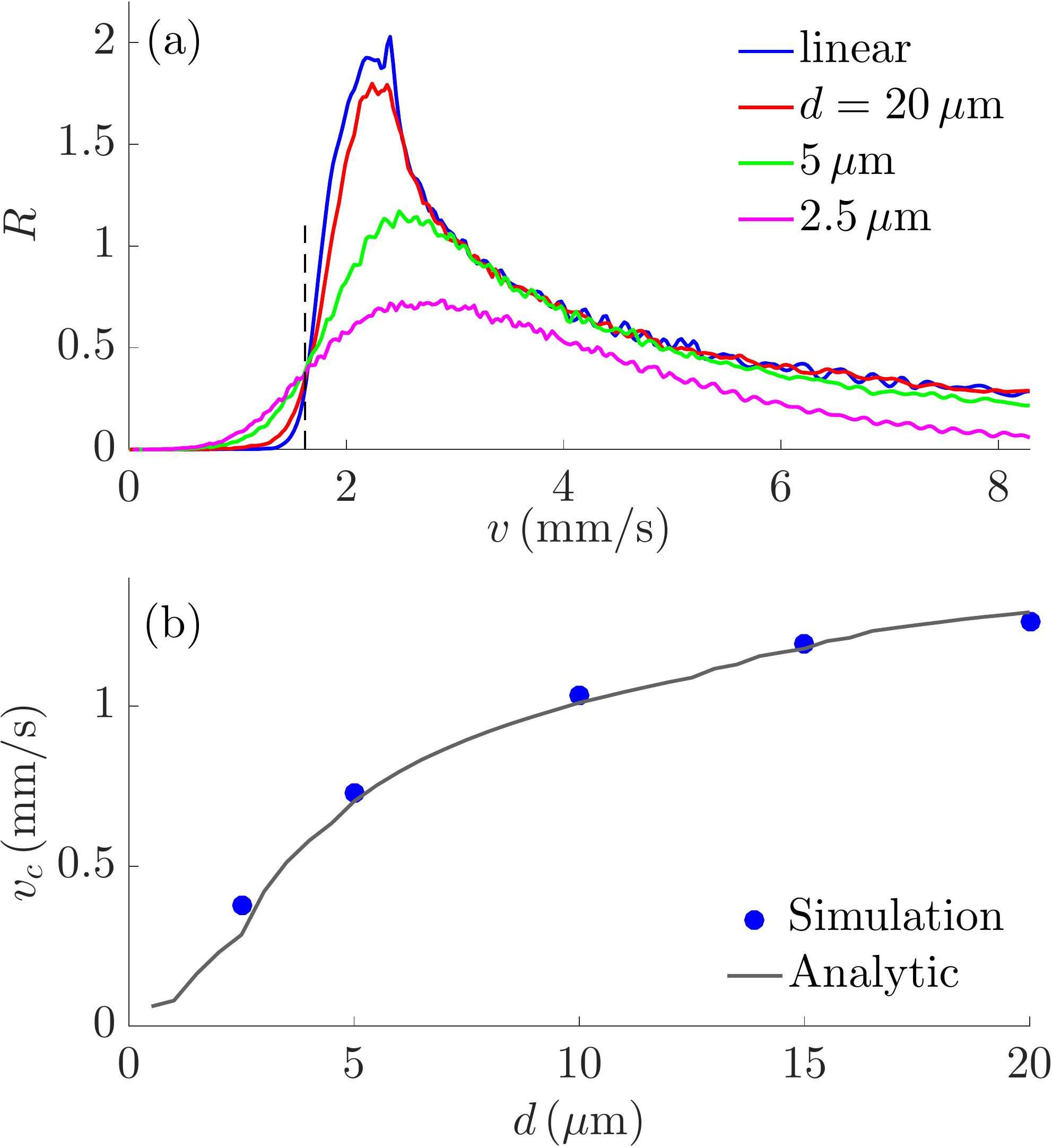}
\caption{(a) Simulated heating rate $R(v)$ for $d=20$, $5$ and $2.5\, \mum$, and their comparison with the heating rate of a linear stirring motion (blue line). 
(b) Critical velocity $v_c$ determined from the simulations (circles) and the Bogoliubov estimate (continuous line) for various values of $d$.  }
\label{Fig3}
\end{figure}
\begin{figure*}[]
\includegraphics[width=1\linewidth]{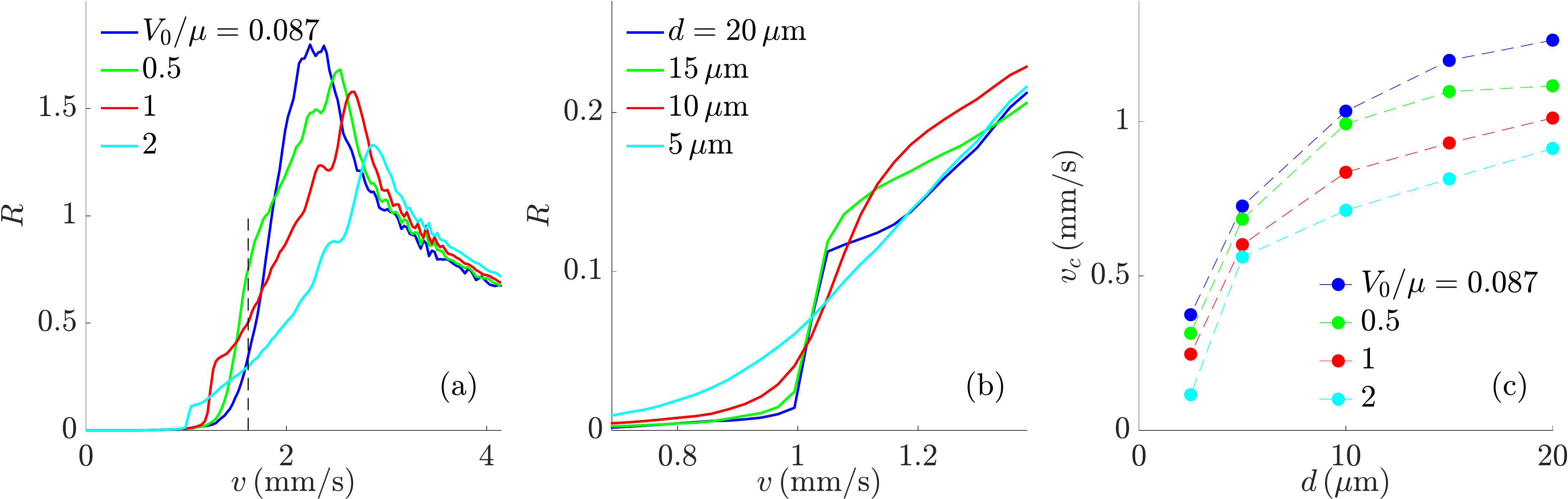}
\caption{(a) $R(v)$ for $d=20\, \mum$ and $V_0/\mu= 0.087$, $0.5$, $1$, and $2$. 
The vertical dashed line denotes the Bogoliubov velocity $v_{\mB}$.
(b) $R(v)$ for $V_0/\mu=2$ and different values of $d$.  
(c) $v_c$ as a function of $d$ for the same $V_0/\mu$ as in panel (a), where the dashed lines are guide to the eye.   }
\label{Fig4}
\end{figure*}

    In Fig. \ref{Fig3}(a) we compare the simulations of linear-oscillating motion with the simulation of a linear motion.  
For a linear stirring motion, we choose $x(t)=vt$ moving at a constant velocity $v$ along the $+x$ direction. 
We determine $R(v)$ and show the results in Fig. \ref{Fig3}(a).
This heating rate gives the onset of dissipation close to $v_\mB$. 
It agrees with the linear-oscillating motion of $d=20\, \mum$ for high $v$ and deviates for $v$ near $v_\mB$.
This deviation is due to the difference between a linear-oscillating and a linear motion, which is more pronounced for small $d$. 
For a large  $d\rightarrow\infty$ our linear-oscillating motion approaches a linear motion and we recover the Landau criterion of superfluidity,  where $v_c$ is approximately equal to $v_\mB$ and dissipation above $v_c$ occurs due to the creation of phonons. 
The oscillating motion results in an additional broadening of the heating rate curve, 
which is responsible for an onset of dissipation at lower velocity below $v_\mB$. 
To quantify this reduction we determine $v_c$ via fitting with the fitting function in Eq. \ref{eqn:fitfunction} and present these results in Fig. \ref{Fig3}(b). 
$v_c$ decreases with decreasing $d$. 
We note that this reduction of $v_c$ is similar to the reduction found for a circular stirring motion in  Ref. \cite{Singh2016}, where $v_c$ was lower for a smaller stirring radius. 
We compare $v_c$ with the estimate of $v_c$ determined using  Eqs. \ref{eq:Rate} and \ref{eqn:fitfunction}. 
This estimate provides an excellent agreement with the simulated $v_c$. 
$v_c$ appears to vanish for $d \rightarrow 0$. 
The dependence of $v_c$ on $d$ is approximately linear for small $d$ and sublinear for large $d$ while approaching $v_c$ of linear motion.

\section{Strong stirring potential} \label{sec:V0st}
\begin{figure}[]
\includegraphics[width=1\linewidth]{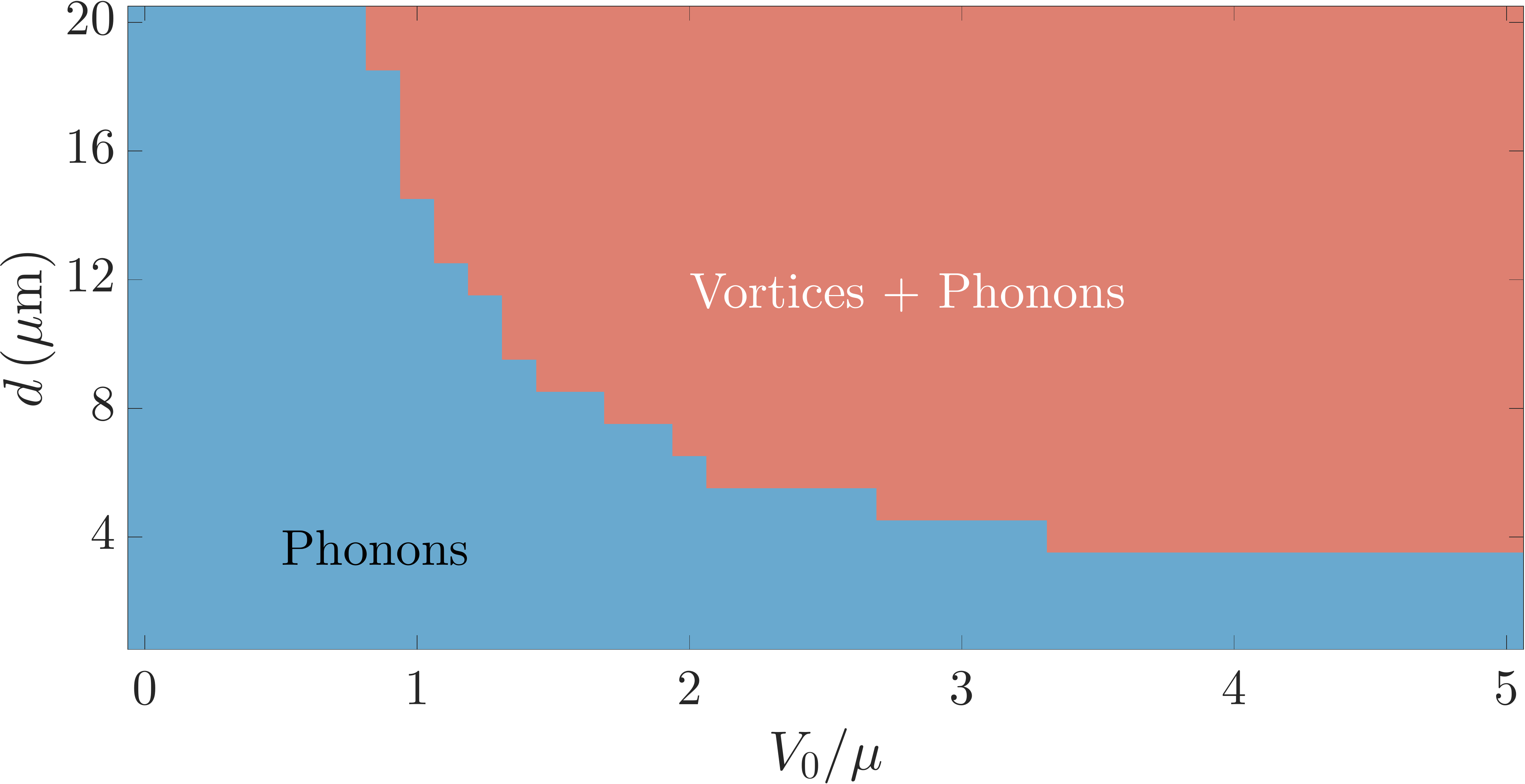}
\caption{Dissipation regimes as a function of $V_0$ and $d$ for stirring velocity $v/v_\text{B} =0.85$ and $T/T_c=0.05$.
The blue and the red shaded area denote the phonon and the vortex regime, respectively.    }
\label{Fig5}
\end{figure}
\begin{figure}[]
\includegraphics[width=1\linewidth]{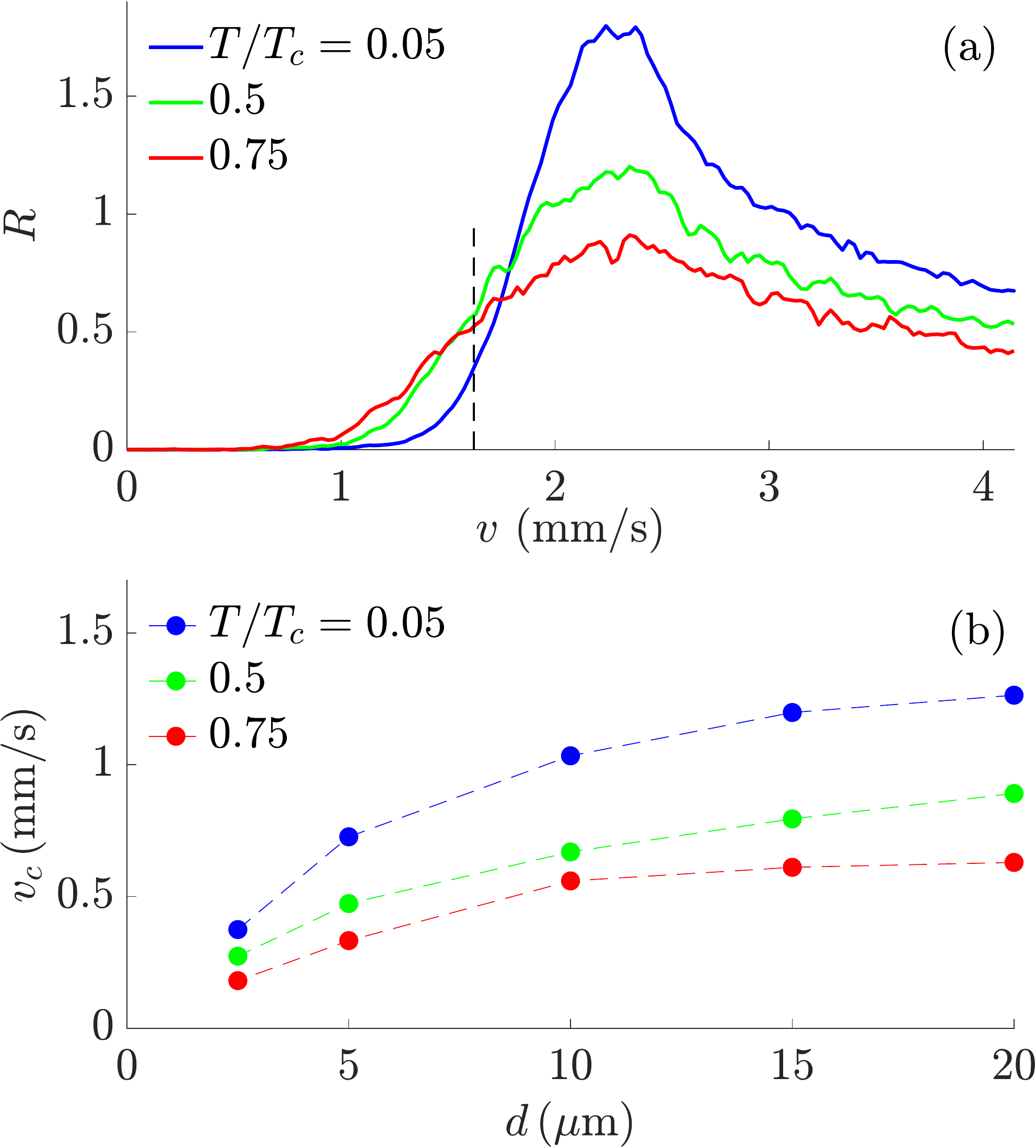}
\caption{(a) $R(v)$ for $d=20\, \mum$ and  $T/T_c=0.05$, $0.5$, and $0.75$. 
(b) $v_c$ as a function of $d$ for various $T/T_c$ as in panel (a).  }
\label{Fig6}
\end{figure}
\begin{figure}[]
\includegraphics[width=1\linewidth]{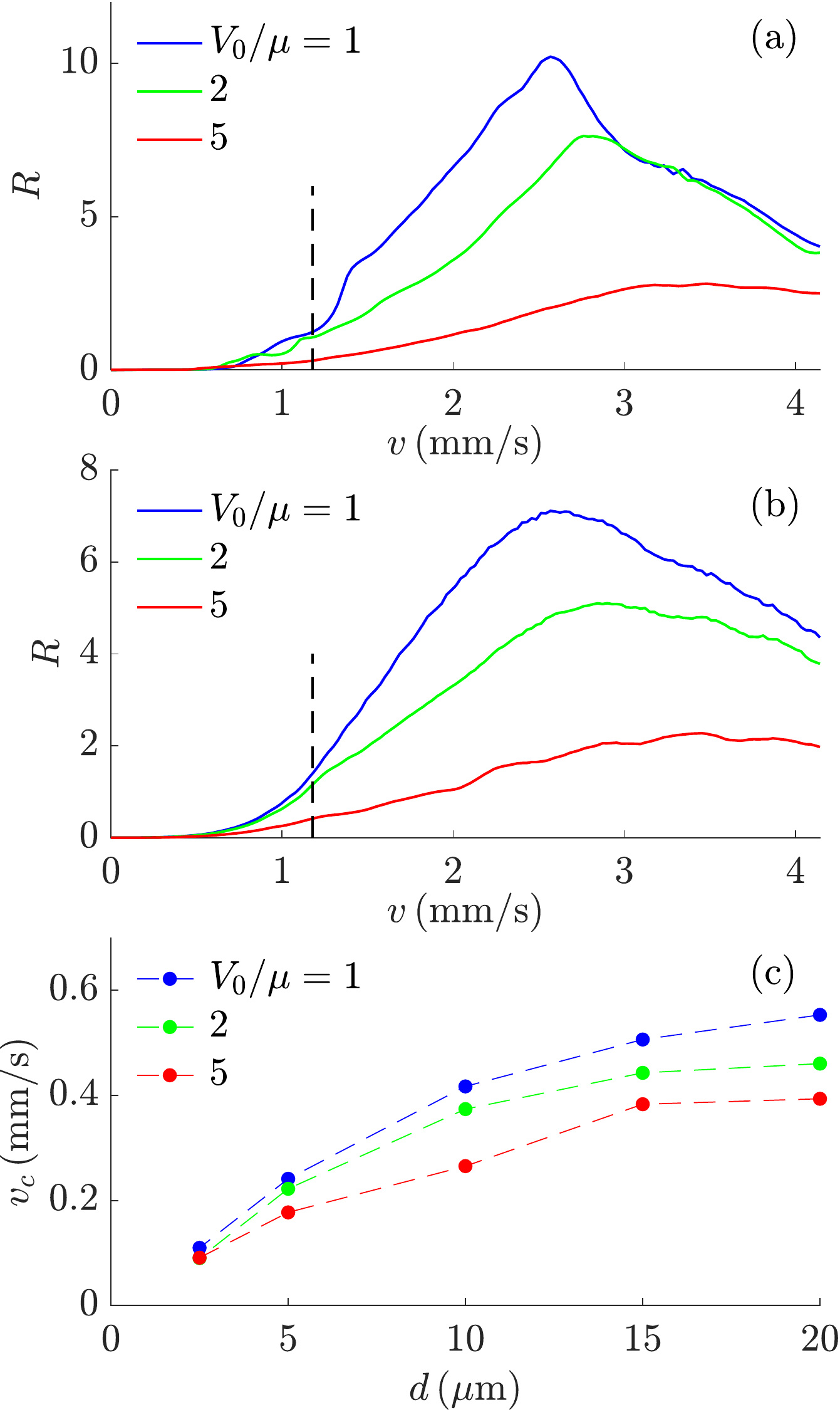}
\caption{(a) Simulated heating rate $R(v)$ of the trapped cloud  for $d=20\, \mum$ and $V_0/\mu=1$, $2$, and $5$.  
(b) $R(v)$ for $d=5\, \mum$ and the same $V_0/\mu$ as in panel (a).  
(c) $v_c$ as a function of $d$ for the same  $V_0/\mu$ as in panel (a).  
The vertical dashed line in (a) and (b) denotes the Bogoliubov velocity $v_{\mB, 0}$; see text.  }
\label{Fig7}
\end{figure}

  We now proceed to examine the influence of a strong stirring potential.
We use the same system parameters as in Sec. \ref{sec:hom} and determine the simulated heating rate $R(v)$ for various $V_0$ and $d$.
In Fig. \ref{Fig4}(a) we show $R(v)$ for $d=20\, \mum$ and $V_0/\mu= 0.087$, $0.5$, $1$ and $2$. 
Compared to a weak stirrer, the heating rate shows qualitative changes for a strong stirrer. 
The heating rate shows a steep increase at the onset of dissipation and then increases slowly with increasing $v$ around $v_\mB$. 
$R(v)$ decreases at high $v$ for all  $V_0$.
For increasing $V_0$, the onset and the maximum of the heating rate shift to a lower and a higher $v$, respectively. 
We show below that this steep onset of heating is associated with the formation of vortex-antivortex pairs. 
We now examine how this steep onset of heating is modified as $d$ is reduced. 
In Fig. \ref{Fig4}(b) we show $R(v)$ for $V_0/\mu=2$ and $d=20$, $15$, $10$, and $5\, \mum$, 
for which the values of $v$ are chosen below $v_\mB$. 
The steep onset  is clearly present for $d=20\, \mum$ as in Fig. \ref{Fig4}(a), which is then washed out for smaller $d$ due to increased broadening of the oscillating motion. There is no steep onset visible at all for $d=5\, \mum$.
This implies suppressed vortex-induced dissipation for small displacements. 
In Fig. \ref{Fig4}(c) we show the results of $v_c$ as a function of $d$ for various $V_0$. 
$v_c$ decreases with decreasing $d$ for all $V_0$, manifesting the reduction of oscillating motion present at all $V_0$.
For high $V_0$ and large $d$, the reduction of $v_c$ with respect to a weak stirrer is mainly due to the creation of vortex pairs, which we demonstrate below.  
For high $V_0$ and small $d$, it is due to both vortex creation and the broadening induced by the oscillating, and therefore accelerated, motion.

     To investigate vortex-induced dissipation we calculate the time evolution of the phase $\phi(x, y)$ of a single plane and a single trajectory during stirring. 
For vortices we calculate the phase winding around the lattice plaquette of size $l \times l$ using $\sum_\Box \delta\phi(x,y)=\delta_x\phi(x,y)+\delta_y\phi(x+l,y)+\delta_x\phi(x+l,y+l)+\delta_y\phi(x,y+l)$, where the phase difference between sites is taken to be $\delta \phi_{x,y}\in(-\pi,\pi]$. 
A phase winding of $+2\pi$ and $-2\pi$ indicates a vortex and an antivortex, respectively. 
We count the total number of vortices and average it over the $z$ direction and  the ensemble.
As an example, we choose a stirring velocity $v/v_\mB= 0.85$ and analyze the time evolution of the vortex number for various $V_0$ and $d$. 
If we find on average at least one vortex pair in the time evolution during stirring, we refer to this as the vortex regime. 
In Fig. \ref{Fig5} we show the dissipation regimes for vortex creation as a function of $V_0$ and $d$.
For large $d$ the onset of vortices sets in for $V_0/\mu \gtrsim 1$.
This vortex region shrinks as $d$ is reduced.
For small $d < 4\, \mum$ the time evolution shows no vortex pair creation for all $V_0$. 
This  explains why a steep onset of heating is present at $d=20$, $15$ and $10\, \mum$, and absent at $d=5 \, \mum$, for $V_0/\mu=2$ in Fig. \ref{Fig4}(b).

\begin{figure*}[]
\includegraphics[width=1\linewidth]{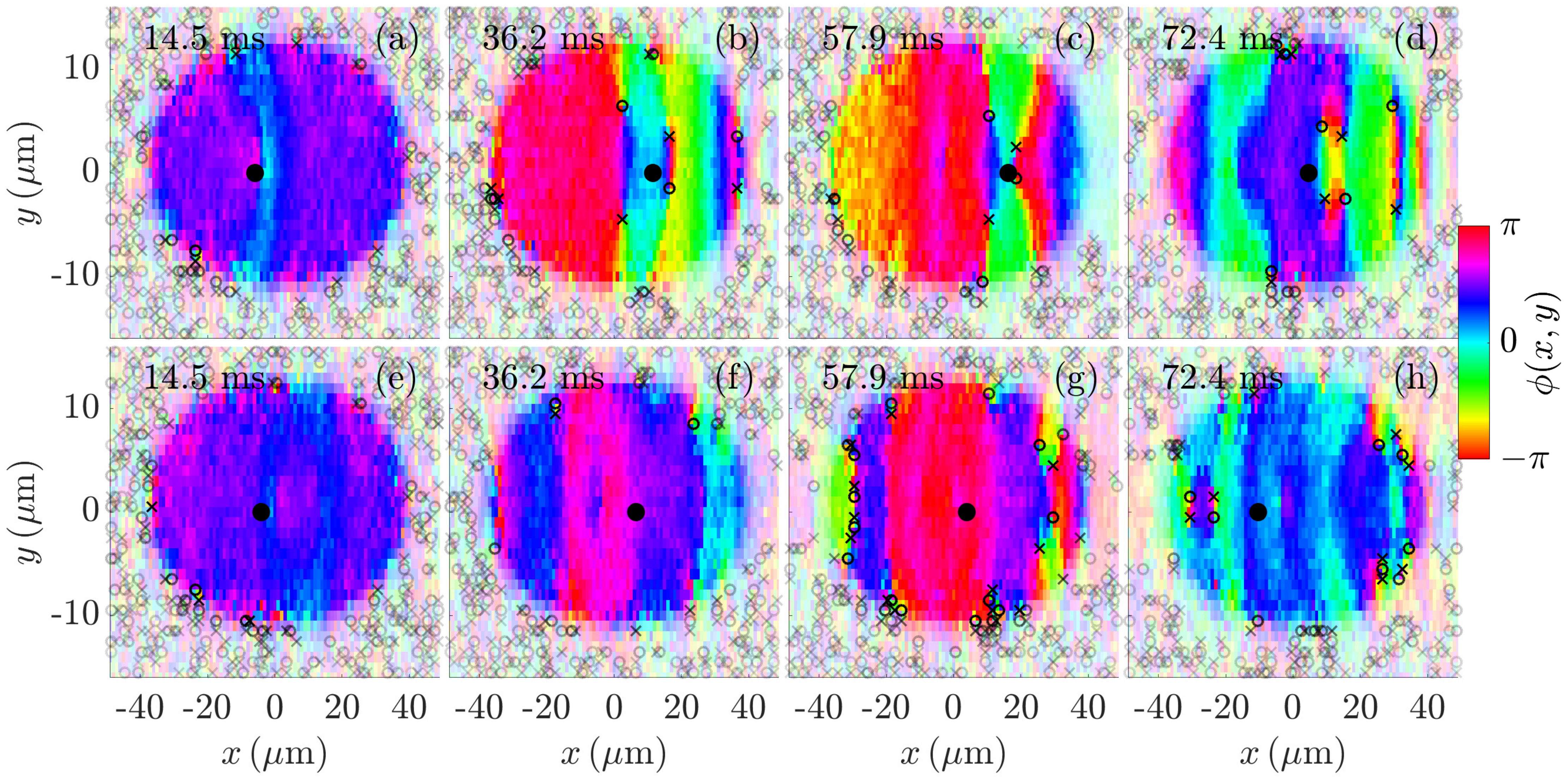}
\caption{Time evolution of the phase $\phi(x, y)$ of the central plane of a single trajectory during stirring at $v=1.1\, \mmms$ and the stirring time $t=14.5$, $36.2$, $57.9$, and $72.4\, \mms$,  for $d=20\, \, \mum$ in panels (a-d) and $d=5\, \, \mum$ in panels (e-h). 
The circles and the crosses represent vortices and antivortices, respectively. The black disc indicates the stirrer moving back and forth along the $x$ direction.
The color transparency outside the Thomas-Fermi radius is chosen according to the density of the thermal cloud. 
  }
\label{FigPhase}
\end{figure*}
\section{High Temperature}\label{sec:temp}

  Next, we examine the influence of temperature. 
For all simulations we use the same density $n$ as above and $V_0/\mu=0.087$.
In Fig. \ref{Fig6}(a) we show $R(v)$ for $d=20\, \mum$ and $T/T_c=0.05$, $0.5$, and $0.75$. 
The broadening and the height of the heating rate curve increases and decreases with increasing $T/T_c$, respectively. 
This results in an onset of dissipation at low velocity for high temperature. 
In Fig. \ref{Fig6}(b) we show $v_c$ as a function of $d$ for the same $T/T_c$ as in Fig. \ref{Fig6}(a). 
$v_c$ decreases with both increasing $T/T_c$ and decreasing $d$. 
While the temperature reduction is more pronounced for large $d$ than for small $d$, the dependence on $d$ is similar at both low and high temperature. 
This temperature reduction of $v_c$ is due to the thermal broadening of the phonon modes and the reduction of the phonon velocity itself at nonzero temperature.

\section{Trapped condensate}\label{sec:trap}

    Finally, we consider a trapped cloud as in the experiments \cite{Ketterle1999}. 
We simulate a cigar-shaped condensate having the central density $n_0= 10.7 \, \mum^{-3}$ and $T/T_c=0.75$. 
The critical temperature $T_c$ is the estimate of a non-interacting trapped system. 
In all simulations we choose $d \leq 25\, \mum$ for stirring the condensate region of the cloud using the same stirring process as above. 
While the cloud size and the temperature are comparable to the experiments,  $n_0$ is smaller by a factor of $10$ than in the experiments to enable our numerical simulations.
We first isolate the effect of oscillating motion and strength of the stirrer in the trapped cloud.
We use $2\sigma=2.0\, \mum$ and calculate the heating rate $R(v)$ for various $d$ and $V_0/\mu$. 
The mean-field energy $\mu= g n_0$ is determined using $n_0$. 
In Fig. \ref{Fig7}(a) we show $R(v)$  for $d=20\, \mum$ and $V_0/\mu=1, \, 2$, and $5$.  
This heating rate shows a behavior that is similar to the case of a homogeneous system at low temperature in Fig. \ref{Fig3}(a). 
Here the strong stirrer response is broadened by the thermal broadening of the phonon modes and the inhomogeneous density of the cloud. 
The onset of heating has contributions of vortex excitations as we demonstrate below. 
This vortex-induced dissipation is suppressed for small $d$ due to increased broadening of the oscillation motion as in Fig. \ref{Fig7}(b).
In Fig. \ref{Fig7}(c) we show $v_c$ and its dependence on $V_0$ and $d$.
In this strong stirrer regime $V_0/\mu \gtrsim 1 $, $v_c$ decreases with both increasing $V_0$ and decreasing $d$. 
This is similar to the case of a homogeneous system and low temperature. 
For large $d$ the reduction in $v_c$ is due to the creation of vortices that occurs at a lower velocity. 
This vortex creation is suppressed for small $d$ due to increased broadening of the oscillating motion.
We compare the results of $v_c$ with the Bogoliubov estimate of the sound velocity in a cigar-shaped cloud. 
A sound wave in the axial direction propagates with velocity $v_\mB= \sqrt{g n_\meff/m}$ near the trap center.  
The effective density $n_\meff=n_0/2$ is determined by averaging the Thomas-Fermi profile in the radial direction \cite{Stringari1998}. 
This results in  $v_{\mB, 0}=1.18\, \mmms$  at the trap center and $v_{\mB,d=20 \, \mu\text{m}}=1.0 \, \mmms$ at $d=20\, \mum$.
The values of $v_c$ for $d=20\, \mum$ are in the range $0.45-0.55 \, \mmms$, which are approximately half of $v_{\mB,d=20 \, \mu\text{m}}$. This reduction of $v_c$ is due to increased thermal fluctuations and the creation of vortices as we show below.

\begin{figure*}[]
\includegraphics[width=1\linewidth]{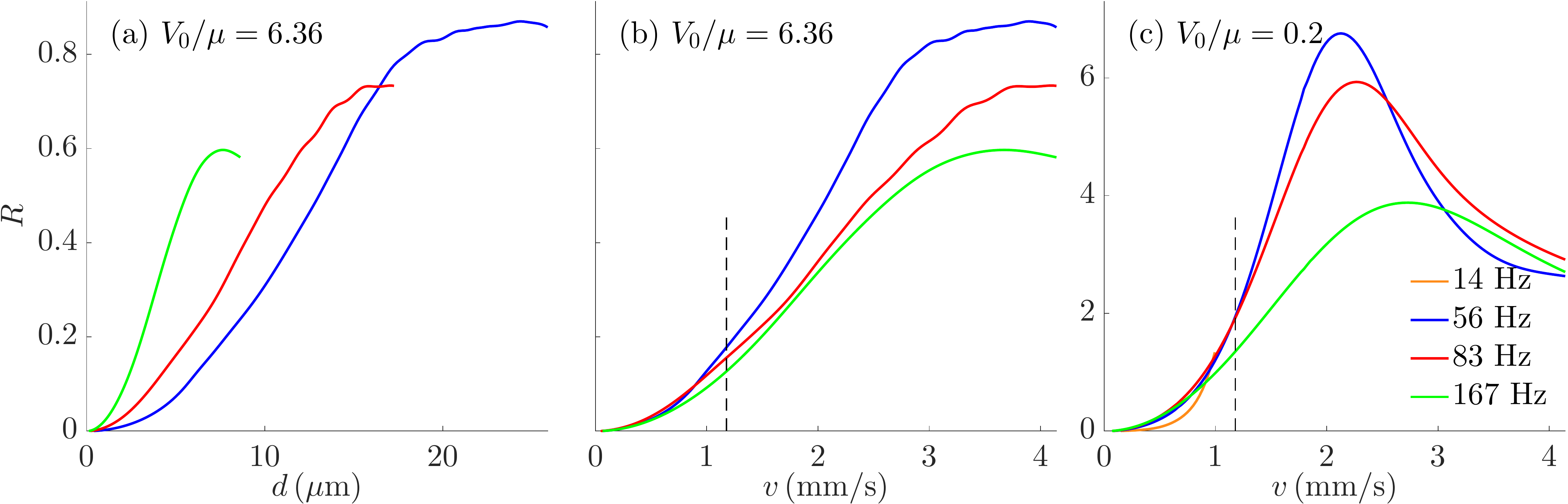}
\caption{(a) $R$ as a function of $d$ for $V_0/\mu =6.36$ and $f=56$, $83$, and $167\, \mHz$. 
(b) $R(v)$ determined using $v=4fd$ and the results of panel (a). 
(c) $R(v)$ for $V_0/\mu =0.2$ and $f=14$, $56$, $83$, and $167\, \mHz$.  }
\label{Fig8}
\end{figure*}

   To understand the role of vortices in the onset of heating we analyze the time evolution of the phase $\phi(x, y)$ of the central plane of a single trajectory belonging to the same system as above. 
We choose the stirring velocity $v=1.1 \, \mmms$, $V_0/\mu=2$, and the same $2\sigma$ as above. 
In Figs. \ref{FigPhase}(a-d) we show $\phi(x, y)$ for $d=20\, \mum$ at the stirring time $t$ of $14.5$, $36.2$, $57.9$, and $72.4\, \mms$.
For $t=14.5\, \mms$ the phase is weakly fluctuating in the condensate cloud, indicating global phase coherence of the system. 
The stirring motion induces fluctuations of the phase of the system. 
For long stirring times these phase fluctuations are increased and result in the creation of vortex pairs. 
This is confirmed by the calculation of the phase winding around each plaquette, following the method described in Sec. \ref{sec:V0st}. 
The resulting phase windings are shown in Fig. \ref{FigPhase}. Since the phase fluctuates strongly in the thermal cloud we observe a vortex plasma outside the Thomas-Fermi region. For $t=14.5\, \mms$ the stirring process creates no vortex pairs in the condensate region. 
For $t=36.2\, \mms$ and longer times, vortex pairs are nucleated from the regions of low density near the stirrer. 
In addition to this vortex shedding, vortex pairs in the thermal cloud travel to low density regions of the condensate cloud. 
In Figs. \ref{FigPhase}(e-h) we show the time evolution of the phase and vortices  for $d=5\, \mum$. 
The vortex shedding mechanism is suppressed as no vortex creation is observed near the stirrer. 
Vortex pairs in the thermal cloud travel to the condensate, which is similar to the case at $d=20\, \mum$.

\begin{figure*}[]
\includegraphics[width=1\linewidth]{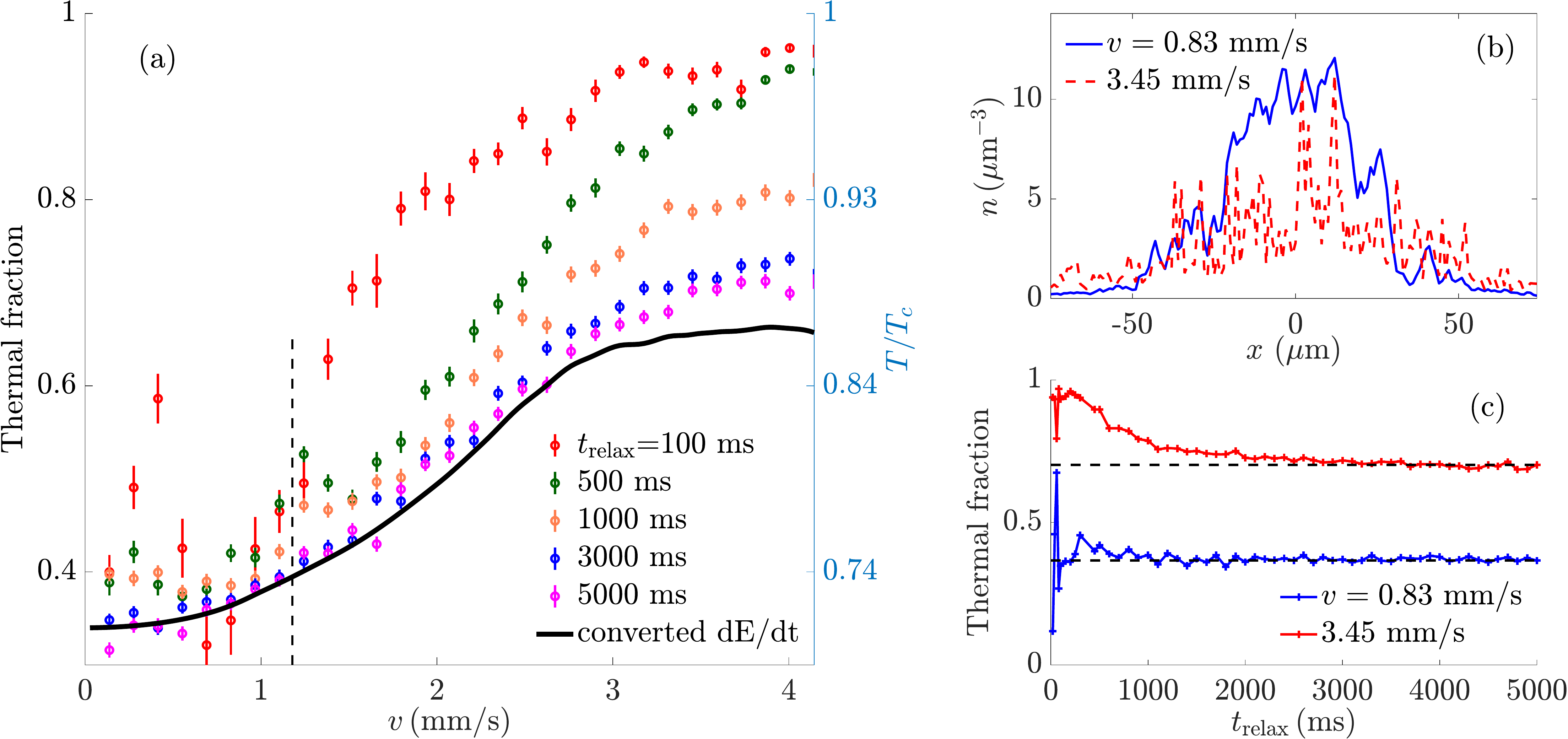}
\caption{(a) Thermal fraction as a function of $v$ for various relaxation times in the range $t_\mrelax=100-5000\, \mms$. 
The black continuous line shows the thermal fraction determined from the heating rate of the system given in Fig. \ref{Fig8}(b).
(b) Density profile $n(x)$ of the cut along the $x$ direction after $t_\mrelax=100\, \mms$ for $v=0.83$ and $3.45\, \mmms$. 
(c) Time evolution of the thermal fraction as a function of $t_\mrelax$ for the same $v$ as in panel (b). }
\label{Fig9}
\end{figure*}

     To compare the simulations with the experiments \cite{Ketterle1999},  we follow the experimental method of stirring, 
 which includes varying $d$ and keeping $f$ fixed for determining the stirring velocity $v=4fd$.
This differs from our method employed above, where we kept $d$ fixed and varied $f$ to determine $v$.
We choose $V_0/ \mu= 6.36$ and $2\sigma= 3.25\, \mum$, which are comparable to the experiments. 
In Fig. \ref{Fig8}(a) we show  $R$ as a function of $d$ for $f=56, \, 83$, and $167\, \mHz$. 
$R(d)$ shows a frequency dependent behavior as expected. 
We show this result as a function of $v=4fd$ in Fig. \ref{Fig8}(b). 
The different frequency results approximately collapse at low velocities, while they differ at intermediate and high velocities. 
$R(v)$ is not suppressed at low velocities and increases continuously with increasing $v$. 
This results in a vanishing $v_c=0$ for the results in Fig. \ref{Fig8}(b).  
The reason for this vanishing $v_c$ is the broadening induced by the oscillating motion, which results in a different velocity dependence in Figs. \ref{Fig8}(b,c). 
The low-velocity regime $v < v_{\text{B},0}$ includes stronger contributions of this broadening for $v$, or equivalently $d$, approaching $0$. 
This vanishing $v_c$ for the comparable parameters to the experiment is in disagreement with the measurement of a nonzero $v_c$ \cite{Ketterle1999}.
As we show below, a nonzero $v_c$ occurs in a non-equilibrium cloud which was used in the experiments, 
providing consistent results with the experiments. 
To rule out a strong stirrer effect, we calculate the heating rate for a weak stirrer of $V_0/ \mu= 0.2$ and various $f$. 
We show these results in Fig. \ref{Fig8}(c).
$R(v)$ increases continuously with increasing $v$ for the same $f$ as in Fig. \ref{Fig8}(b) and hence the critical velocity is zero. 
In this weak stirrer regime we recover the maximum of the heating rate, which was broadened by strong $V_0$ in Fig. \ref{Fig8}(b). 
We also show $R(v)$ determined at a lower frequency of $f=14\, \mHz$ and $V_0/ \mu= 0.2$ in Fig. \ref{Fig8}(c).
This result shows suppressed heating at low velocities since the oscillating motion includes larger $d$ and the resulting broadening is reduced. 
This yields a critical velocity of $v_c=0.32 \, \mmms$, contrary to the results of high $f$ in Figs. \ref{Fig8}(b, c).

   To understand the observation of a non-zero $v_c$ in the experiment, we expand on our analysis for determining $v_c$ from the thermal fraction of the stirred cloud rather than the analysis based on the total energy $E = \langle H_0 \rangle$, as up to now. 
The measured $v_c$ is determined by analyzing the thermal fraction ($1- N_0/N$) as a function of the stirring velocity $v=4fd$. 
We extract the condensate fraction $N_0/N$ from the central region of the in situ density profile after $t_{\mrelax}=100\, \mms$ relaxation time, such as in Fig. \ref{Fig9}(b), via fitting it with a Thomas-Fermi profile.  
This also enables us to determine the relative temperature $T/T_c$ via $T/T_c = (1-N_0/N)^{1/3}$ \cite{Pethick2008}. 
In Fig. \ref{Fig9}(a) we show the thermal fraction as a function of $v$ for $f=56\, \mHz$ and various relaxation times in the range $t_{\mrelax}=100- 5000\, \mms$. 
The results show a dependence on the relaxation time, implying that the system is not relaxed at short $t_{\mrelax}$ as in the experiment.
For $t_{\mrelax}=100\, \mms$, the thermal fraction response is similar to the measurements and yields a critical velocity of $v_c= 0.73 \pm 0.44\, \mmms$. 
The high uncertainty of $v_c$ is a consequence of the non-equilibrium behavior of the stirred cloud. 
In Fig. \ref{Fig9}(b) we show the density distribution of the cloud after  $t_{\mrelax}=100\, \mms$ for $v=0.83$ and $3.45\, \mmms$.
Both low and high velocity density profiles are not equilibrated and differ substantially from a thermalized cloud, due to the excitations induced by stirring.  
This is confirmed by the time evolution of the extracted thermal fraction shown as a function of $t_{\mrelax}$ in Fig. \ref{Fig9}(c). 
The thermal fraction fluctuates at short $t_{\mrelax}$ and then slowly relaxes to the equilibrium value for $t_{\mrelax}$ of about $3-5\, \ms$.
We elaborate on this in Sec. \ref{sec:relax}.
This slow relaxation is also reflected in the thermal fraction shown as a function of $v$ in Fig. \ref{Fig9}(a), 
where the result converges as $t_{\mrelax}$ approaches $3\, \ms$ and $5\, \ms$. 
In Fig. \ref{Fig9}(a) we also show the thermal fraction obtained from the heating rate based on the total energy of the system for $f=56\, \mHz$ in Fig. \ref{Fig8}(b). 
For this, we numerically invert the energy change of the temperature of the equilibrium system, see also Ref. \cite{Singh2017}.
This result agrees with the thermal fraction determined after $t_{\mrelax}=3\, \ms$ and $5\, \ms$. 
This confirms that the thermal fraction of the equilibrium system increases continuously at low stirring velocities and hence no critical velocity is observed, as in the case of results in Fig.  \ref{Fig8}(b).

\begin{figure*}[]
\includegraphics[width=1\linewidth]{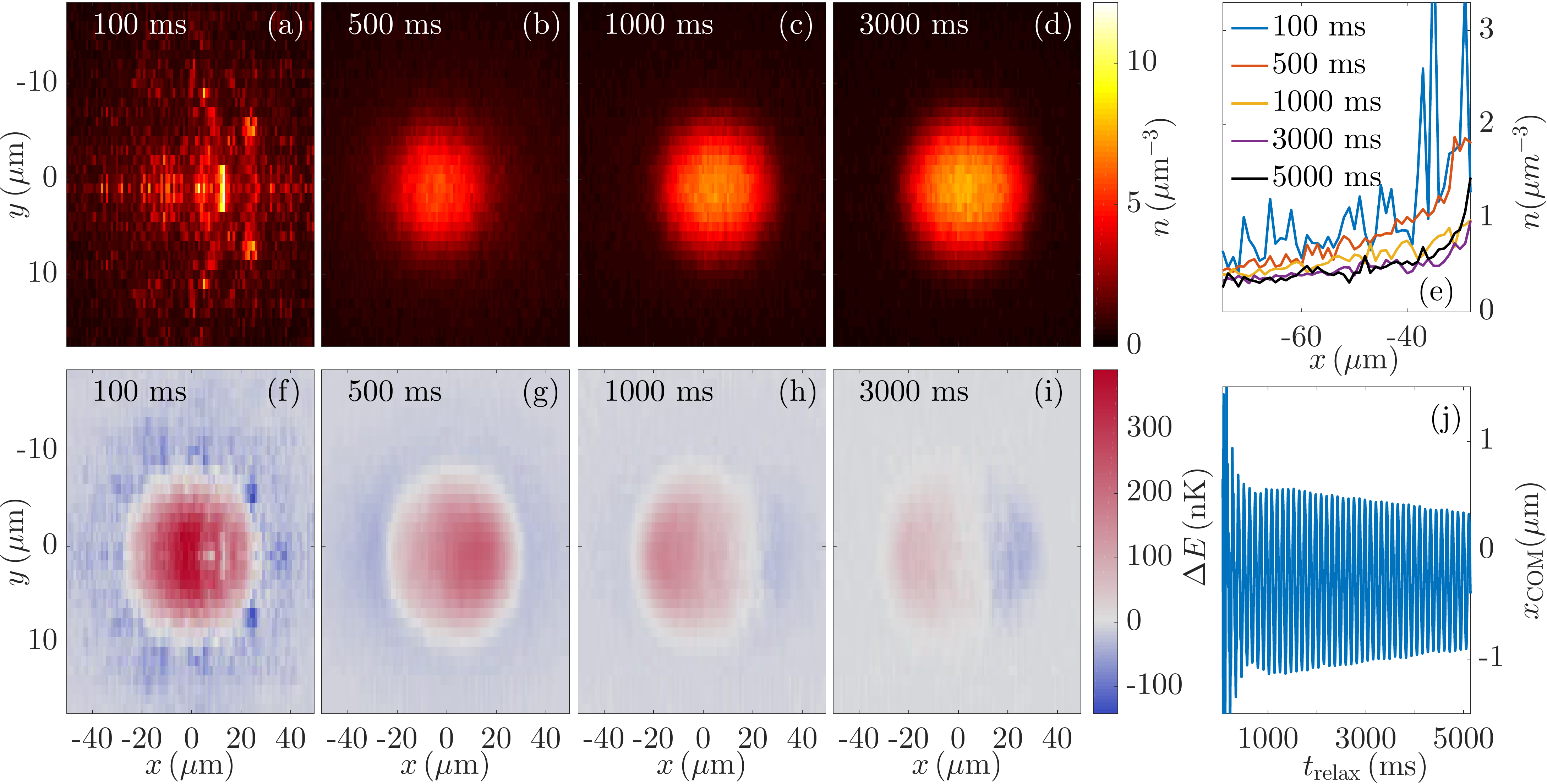}
\caption{(a-d) Recondensation dynamics of the density profile $n(x, y)$ for the relaxation times: $t_\mrelax= 100$, $500$, $1000$, and $3000\, \mms$. 
The cloud was stirred using $v=3.45\, \mmms$ and $f=56\, \mHz$. 
(e) Density cut $n(x)$ along the $x$ direction showing the thermal wing of the same cloud as in panels (a-d).
 $n(x)$ of a nearly equilibrated system at $t_\mrelax= 5000\, \mms$ is also shown. 
(f-i) Time evolution of the energy difference $\Delta E (x, y) = [E_{t_\mrelax} (x, y) - E_f (x, y) ]/n_0$ for the same system as in panels  (a-d).  
$E_{t_\mrelax} (x, y) $ is the energy at time $t_\mrelax$ and $E_f (x, y)$ is the final energy of the equilibrated system. 
(j) Center of mass motion $x_\mCOM(t)$ of the corresponding system as a function of $t_\mrelax$.
}
\label{Fig10}
\end{figure*}
\section{Relaxation Dynamics}\label{sec:relax}

  To understand the relaxation dynamics  we analyze the time evolution of the density and the energy of the system after stirring. 
We use the same system and stirrer parameters as above and stir the cloud using  $v=3.45\, \mmms$ and $f=56\, \mHz$. 
We calculate the density profile $n(x, y)$ of the central plane and average it over the ensemble. 
In Figs. \ref{Fig10}(a-d) we show the time evolution of $n(x, y)$ after stirring for $t_\mrelax= 100$, $500$, $1000$, and $3000\, \mms$. 
For $t_\mrelax= 100\, \mms$ the density distribution shows strong fluctuations that are notably distinct from a Thomas-Fermi profile, for this choice of the stirring parameters, which are motivated by the experiments \cite{Ketterle1999}. 
Since the stirring potential is large and the stirring velocity is above the phonon velocity,  
strong depletion of the condensate region is observed for $t_\mrelax= 100\, \mms$. 
The density distribution shows slow recondensation dynamics and we recover a Thomas-Fermi distribution for $t_\mrelax= 500\, \mms$. 
The density of the condensate region increases and finally reaches the equilibrium value for $t_\mrelax \sim 3000\, \mms$. 
In Fig. \ref{Fig10}(e) we show the thermal wing of the central cut along the $x$ direction of the density profiles in Figs. \ref{Fig10}(a-d).
The thermal wing is strongly fluctuating at short $t_\mrelax$ and then reaches the equilibrium value for  $t_\mrelax \sim 3\, \ms$. 
We also show the result at $t_\mrelax=5\, \ms$, which supports $t_\mrelax$ to be about $3-5\, \ms$.
This slow relaxation is the origin of the higher thermal fraction than the equilibrium result in Fig. \ref{Fig9}(a).

     We now analyze the energy flow dynamics of the stirred cloud. 
We calculate the local energy difference per particle, i.e., $\Delta E (x, y) = [E_{t_\mrelax} (x, y) - E_f (x, y) ]/n_0$, where $E_{t_\mrelax} (x, y)$ is the energy at time $t_\mrelax$ and $E_f (x, y)$ is the final energy of the equilibrium system. 
Both  $E_{t_\mrelax} (x, y) $ and $E_f (x, y)$ are integrated over the $z$ direction and averaged over the ensemble. 
Specifically, $E(x, y)$ is obtained via
\begin{equation}
E(x, y) = \braket{ H_\mkin(x,y) } + \braket{ H_\mpot(x,y) } + \braket{ H_\mint(x,y) },
\end{equation}
with
\begin{align}
\braket{ H_\mkin(x,y) } &=- \frac{J}{2} \sum_{\br^\prime, z} \braket{ \psi^\ast(\br) \psi(\br^\prime) +  \psi(\br) \psi^\ast(\br^\prime)} \\
\braket{ H_\mpot(x,y) } &= \sum_z V(\br) \braket{ |\psi(\br)|^2 } \\
\braket{ H_\mint(x,y) } &=  \frac{U}{2} \sum_z  \braket{ |\psi(\br)|^4  },
\end{align}
where $\br^\prime$ represents the neighboring sites of $\br = (x, y, z)$. 
$V(\br)$ is the harmonic trapping potential. $U=g/l^3$ and $J=\hbar^2/(2ml^2)$ are the Bose-Hubbard parameters, where $l$ is the discretization length. 
In Figs. \ref{Fig10}(f-i) we show $\Delta E (x, y)$ corresponding to the results presented in Figs. \ref{Fig10}(a-d).
The stirred cloud reveals slow energy flow dynamics between the condensate and the thermal cloud of the system. 
For $t_\mrelax= 100\, \mms$ most of the stirring-induced excess energy lies within the condensate region. 
The system then slowly relaxes by transporting this excess energy to the thermal wings of the cloud and finally achieves equilibration for $t_\mrelax \sim 3000\, \mms$. 
The energy dynamics also indicates a sloshing motion that is associated with the dipole mode of the trap, which was excited by the stirring process even though the stirring frequency was detuned above the axial trap frequency. 
To investigate this dipole motion we calculate the center-of-mass (COM) of the cloud by 
$x_\mCOM= \sum_{i=1}^{N_l} n_i x_i/  \sum_{i=1}^{N_l} n_i$, where $N_l$ is the number of lattice sites and $n_i$ is the density at site $i=(x_i, y_i, z_i)$. 
We average $x_\mCOM$ over the radial direction and the ensemble. 
In Fig. \ref{Fig10}(j) we show the time evolution of $x_\mCOM(t)$ as a function of $t_\mrelax$. 
$x_\mCOM(t)$ oscillates at the axial trap frequency as expected. 
After stirring,  $x_\mCOM(t)$ shows a fast decay initially and then a slow relaxation occurring above $5 \, \ms$. 
The energy of this dipole mode $E_\mCOM= m \omega^2 x_\mCOM^2/2$ is approximately $0.004\, \mnK$, 
which is small compared to the energy induced by the stirring process $\Delta E=23.57 \, \mnK$.
However, for low stirring velocities this energy becomes comparable to the stirring-induced energy and the relaxation of this dipole mode essentially governs the relaxation dynamics of the system. This is responsible for the fluctuations of the thermal fraction at low stirring velocities in  Fig. \ref{Fig9}(a).

\section{Conclusions}\label{sec:con}
In conclusion, we have determined the superfluid behavior of a cigar-shaped condensate by stirring it with a repulsive Gaussian potential oscillating back and forth along the axial direction. 
Using both classical-field simulations and perturbation theory, we have analyzed the induced heating rate, based on the total energy of the system, as a function of the stirring velocity $v=4fd$, where $f$ is the frequency and $d$ is the displacement. 
As the key result, we have shown that the onset of dissipation, identified via a critical velocity $v_c$, is influenced by the oscillating motion, the strength of the potential, the temperature and the inhomogeneous density of the cloud. 
The range of these features that we study include the regime of large stirring potential and high temperature, that are similar to the experimental parameters of Ref. \cite{Ketterle1999}. For this regime we find a vanishing critical velocity, if the heating rate is based on the total energy. 
In contrast to this result, we find a non-zero critical velocity if the heating rate is based on the thermal fraction of the cloud after $100\, \mms$ relaxation, that is consistent with the  experimental findings.
This discrepancy derives from the slow relaxation of the cloud, which is significantly out-of-equilibrium at $100\, \mms$.
We find that the relaxation occurs on a timescale of $5\, \ms$. 
Furthermore, we have mapped out the regimes of vortex- and phonon-induced dissipation, and have identified a long-lived dipole oscillation contributing to the slow relaxation of the cloud.

The central points of our study are system features and phenomena in cold atom clouds that apply to any study of superfluidity in these systems and have relevance to the studies in polariton condensates \cite{Alberto2009, Giovanni2017} and photonic fluids \cite{David2016, Claire2018, Huynh2021}. 
As such, our study supports the further exploration and understanding of superfluidity.

\section*{acknowledgements}
This work was supported by the Deutsche Forschungsgemeinschaft (DFG) in the framework of SFB 925 – project ID 170620586 
and the excellence cluster  `Advanced Imaging of Matter’ - EXC 2056 - project ID 390715994.
V.P.S. acknowledges funding by the Cluster of Excellence ‘QuantumFrontiers’ - EXC 2123 - project ID 390837967.

%-----------Appendix--------

\appendix

\section{Bogoliubov theory of the Bose gas}
Within second quantization the Hamiltonian of a weakly interacting Bose gas in momentum space is given by
\begin{equation}
    \hat{H}_0 = \sum_{\bk} \epsilon_k \hat{a}_{\bk}^\dagger \hat{a}_{\bk} + \frac{g}{2 V} \sum_{\bk,\bp,\bq} \hat{a}_{\bk+\bm{p}}^\dagger \hat{a}_{\bq-\bp}^\dagger \hat{a}_{\bq} \hat{a}_{\bk},
\end{equation}
where $V$ is the volume and $\epsilon_k=\frac{\hbar^2 k^2}{2 m}$. At $T=0$ we may assume macrocopic occupation of the ground state. Then we may diagonalize this Hamiltonian with a Bogoliubov transformation $\hat{b}_\bk=u_k \hat{a}_\bk-v_k \hat{a}_{-\bk}^\dagger$ where $u_k^2=(\hbar \omega_k + \epsilon_k + g n)/(2 \hbar \omega_k)$ and $v_k^2=(-\hbar \omega_k + \epsilon_k + g n)/(2 \hbar \omega_k)$. The diagonalized Hamiltonian is approximately
\begin{equation}
    \hat{H}_0 = \sum_{\bm{k}} \hbar \omega_k \hat{b}_{\bm{k}}^\dagger \hat{b}_{\bm{k}},
\end{equation}
where $\hbar \omega_k = \sqrt{\epsilon_k (\epsilon_k + 2 m v_B^2)}$.

\section{Analytic heating rate}\label{sec:appRate}
Let us expand the periodic motion into a Fourier series of the form
\begin{equation}
    x(t)=\sum_{\lambda=0}^{N} a_\lambda \cos(\omega_\lambda t).
\end{equation}
We write the Fourier transform of the stirring potential as
\begin{equation}
    V_{\bk}(t) = V_{\bk} \prod_{\lambda=0}^{N} e^{i [k_x a_\lambda \cos(\omega_\lambda t)]},
\end{equation}
where $V_{\bk}=\frac{2 \pi V_0 \sigma^2}{A} \delta_{k_z} e^{-k^2 \sigma^2/2} e^{i k_y y_0}$ is a time in-dependant prefactor. We substitute $\mathcal{A}_{\bk}=-\frac{i}{\hbar} (u_k+v_k) \sqrt{N_0}$ to write the equation of motion in the simple form
\begin{equation}
    \partial_t A_{\bk}(t) = V_{\bk} \mathcal{A}_{\bk} e^{-i \omega_k t} \prod_{\lambda=0}^{N} e^{i [k_x a_\lambda \cos(\omega_\lambda t)]}.
\end{equation}
To integrate this expression we first use the Jacobi-Anger expansion
\begin{equation}
    e^{i [k_x a_\lambda \cos(\omega_\lambda t)]} = \sum_{\nu=-\infty}^\infty i^\nu J_\nu(k_x a_\lambda) e^{i \nu \omega_\lambda t},
\end{equation}
to get
\begin{equation}
    \partial_t A_{\bk}(t) = V_{\bk} \mathcal{A}_{\bk} e^{-i \omega_k t} \prod_{\lambda=0}^{N} \sum_{\nu=-\infty}^\infty i^\nu J_\nu(k_x a_\lambda) e^{i \nu \omega_\lambda t}.
\end{equation}
To carry out the time integration we combine the exponentials
\begin{multline}
    \partial_t A_{\bk}(t) = V_{\bk} \mathcal{A}_{\bk} \smashoperator[l]{\sum_{\nu_0,\dots,\nu_N=-\infty}^\infty} \prod_{\lambda=0}^N \left[i^{\nu_\lambda}J_{\nu_\lambda}(k_x a_\lambda)\right]\\ e^{i (\omega_{eff}-\omega_k) t},
\end{multline}
where we substituted $\omega_{eff} = \sum_{\lambda=0}^N \nu_\lambda \omega_\lambda$. 
We employ the integral
\begin{equation}
    \int_0^t \diff{t} e^{i (\omega_{eff}-\omega_k) t} = 2 e^{-i(\omega_k-\omega_{eff}) t/2} \frac{\sin[(\omega_k-\omega_{eff}) t/2]}{\omega_k-\omega_{eff}}.
\end{equation}
Thus the solution for $A_{\bk}(t)$ is
\begin{multline}
    A_{\bk}(t) =2 V_{\bk} \mathcal{A}_{\bk} \smashoperator[l]{\sum_{\nu_0,\dots,\nu_N=-\infty}^\infty} \prod_{\lambda=0}^N \left[i^{\nu_\lambda}J_{\nu_\lambda}(k_x a_\lambda)\right] e^{-i(\omega_k-\omega_{eff}) t/2}\\ \frac{\sin[(\omega_k-\omega_{eff}) t/2]}{\omega_k-\omega_{eff}}.
\end{multline}
Taking the absolute square we obtain
\begin{multline}
    |A_{\bk}(t)|^2 = 4 |V_{\bk}|^2 |\mathcal{A}_{\bk}|^2 \smashoperator[l]{\sum_{\nu_0,\nu'_0,\dots,\nu_N,\nu'_N=-\infty}^\infty} \prod_{\lambda=0}^N \left[i^{\nu_\lambda-\nu'_\lambda} \right]\\ \times\prod_{\lambda=0}^N \left[J_{\nu_\lambda}(k_x a_\lambda)J_{\nu'_\lambda}(k_x a_\lambda)\right]\\ \times\frac{\sin[(\omega_k-\omega_{eff}) t/2]}{\omega_k-\omega_{eff}} \frac{\sin[(\omega_k-\omega'_{eff}) t/2]}{\omega_k-\omega'_{eff}},
\end{multline}
where $\omega'_{eff} = \sum_{\lambda=0}^N \nu'_\lambda \omega_\lambda$. We approximate this expression by only considering terms with $\omega_{eff}=\omega'_{eff}$. This restriction reduces the number of summation indices by 1. We drop the sum over $\nu'_0$ and write $\nu'_0=\nu_0+\sum_{\xi=1}^N (1+2\xi) (\nu_\xi-\nu'_\xi)$. Thus we get
\begin{multline}
    |A_{\bk}(t)|^2 = 4 |V_{\bk}|^2 |\mathcal{A}_{\bk}|^2 \smashoperator[lr]{\sum_{\nu_0,\nu_1,\nu'_1,\dots,\nu_N,\nu'_N=-\infty}^\infty} (-1)^{\sum_{m=1}^{N} m(\nu_m-\nu'_m)}\\ \times J_{\nu_0}(k_x a_0) J_{\nu_0+\sum_{\xi=1}^N (1+2\xi) (\nu_\xi-\nu'_\xi)}(k_x a_0)\\ \times\prod_{\lambda=1}^N \left[J_{\nu_\lambda}(k_x a_\lambda)J_{\nu'_\lambda}(k_x a_\lambda)\right]\frac{\sin^2[(\omega_k-\omega_{eff}) t/2]}{(\omega_k-\omega_{eff})^2}.
\end{multline}
We define the function
\begin{multline}
    B(k_x)=J_{\nu_0}(k_x a_0) J_{\nu_0+\sum\limits_{\xi=1}^N (1+2\xi) (\nu_\xi-\nu'_\xi)}(k_x a_0)\\ \times\prod_{m=1}^N (-1)^{m(\nu_m-\nu'_m)} J_{\nu_m}(k_x a_m)J_{\nu'_m}(k_x a_m)
\end{multline}
and write the energy change in the system as
\begin{align}
  \braket{\Delta E(t)} &= \sum_{\bk} \hbar \omega_k 4 |V_{\bk}|^2 |\mathcal{A}_{\bk}|^2\nonumber  \\
 &\quad   \times  \smashoperator[lr]{\sum_{\nu_0,\nu_1,\nu'_1,\dots,\nu_N,\nu'_N=-\infty}^\infty} B(k_x) \frac{\sin^2[(\omega_k-\omega_{eff}) t/2]}{(\omega_k-\omega_{eff})^2}.
\end{align}
Using the approximation $\frac{\sin(\alpha t/2)^2}{\alpha^2} = \frac{\pi}{2}t \delta(\alpha)$ we obtain an expression for the heating rate
\begin{equation}
    \frac{\diff{E}}{\diff{t}} = \sum_{\bk}2 \pi \hbar \omega_k |V_{\bk}|^2 |\mathcal{A}_{\bk}|^2 \smashoperator[lr]{\sum_{\nu_0,\nu_1,\nu'_1,\dots,\nu_N,\nu'_N=-\infty}^\infty} B(k_x) \delta(\omega_k-\omega_{eff})
\end{equation}
We convert the sum to an integral $\sum_{k_x,k_y}=\frac{A}{(2 \pi)^2}\int \diff{k_x} \diff{k_y}$, where $A$ is the area (x-y-plane) of the system. Furthermore, we simplify $2 \pi \hbar \omega_k |V_{\bk}|^2 |\mathcal{A}_{\bk}|^2=\frac{4 \pi^3 V_0^2 \sigma^4 N_0}{m A^2} \delta_{k_z} k^2 e^{-k^2 \sigma^2} $, where we used $(u_k+v_k)^2 \omega_k=\frac{\hbar k^2}{2m}$. Thus we are left with
\begin{equation}
    R =\frac{\hbar \sigma^2}{m} \int \diff{k_x} \diff{k_y} k^2 e^{-k^2 \sigma^2}  \smashoperator[lr]{\sum_{\nu_0,\nu_1,\nu'_1,\dots,\nu_N,\nu'_N=-\infty}^\infty} B(k_x) \delta(\omega_k-\omega_{eff}).
\end{equation}
We choose to scale the heating rate as $R=\hbar\frac{\diff{E}}{\diff{t}}/(V_0^2 N_{\text{cyl}})$ where $N_{\text{cyl}}=\pi N \sigma^2/A$. The Bogoliubov spectrum is $\omega_k=\hbar^{-1} \sqrt{\frac{\hbar^2 k^2}{2 m} (\frac{\hbar^2 k^2}{2 m}+2 m v_B^2)}$, where $v_B$ is the sound velocity. To simplify the following calculation we define $\tilde{v}_B=v_B m \sigma/ \hbar$, $\tilde{k}_i = k_i \sigma$, $\tilde{\omega}_i=\omega_i \frac{m \sigma^2}{\hbar}$ as dimensionless quantities. Thus the scaled heating rate is
\begin{equation}
    R =\int \diff{\tilde{k}_x} \diff{\tilde{k}_y} \tilde{k}^2 e^{-\tilde{k}^2}  \smashoperator[lr]{\sum_{\nu_0,\nu_1,\nu'_1,\dots,\nu_N,\nu'_N=-\infty}^\infty} B(\frac{\tilde{k}_x}{\sigma}) \delta(\tilde{\omega}_k-\tilde{\omega}_{eff}),
\end{equation}
where we used a property of the delta distribution $\delta(\frac{\hbar}{m\sigma^2}(\tilde{\omega}_k-\tilde{\omega}_{eff}))=\frac{m \sigma^2}{\hbar} \delta(\tilde{\omega}_k-\tilde{\omega}_{eff})$. In these units $\tilde{\omega}_k=\sqrt{\frac{\tilde{k}^2}{2} (\frac{\tilde{k}^2}{2}+2 \tilde{v}_B^2)}$, $\tilde{\omega}_{eff} = \sum_{\lambda=0}^N \nu_\lambda \tilde{\omega}_\lambda$ and $\tilde{\omega}_\lambda=\frac{m \sigma^2}{\hbar}\omega_\lambda$.

The zeros of the delta function lie at
\begin{equation}
    \tilde{k}_c^2 = -2 \tilde{v}_B^2 + \sqrt{4 \tilde{v}_B^4+4 \tilde{\omega}_{eff}^2}.
\end{equation}
We use the delta to solve the $\tilde{k}_y$ integral. The zeros are at
\begin{equation}
    \tilde{k}_{y,0} = (-2 \tilde{v}_B^2 - \tilde{k}_x^2 + \sqrt{4 \tilde{v}_B^4+4 \tilde{\omega}_{eff}^2})^{1/2}.
\end{equation}
Using the property $\delta(f(x))=\frac{\delta(x-x_0)}{|f'(x_0)|}$ and the identity $\tilde{k}_c^2=\tilde{k}_x^2+\tilde{k}_{y,0}^2$ we obtain
\begin{equation}
    R = \smashoperator[lr]{\sum_{\nu_0,\nu_1,\nu'_1,\dots,\nu_N,\nu'_N=-\infty}^\infty} \tilde{k}_c^2 e^{-\tilde{k}_c^2} \int_{-\tilde{k}_c}^{\tilde{k}_c} \diff{\tilde{k}_x} B(\frac{\tilde{k}_x}{\sigma}) (\frac{\partial \tilde{\omega_k}}{\partial k_y}|_{\tilde{k}_y=\tilde{k}_{y,0}})^{-1}.
\end{equation}
Finally, we solve the derivative and substitute $\kappa=\frac{\tilde{k}_x}{\tilde{k}_c}$ to get
\begin{equation}
    R=\smashoperator[lr]{\sum_{\nu_0,\nu_1,\nu'_1,\dots,\nu_N,\nu'_N=-\infty}^\infty}  \tilde{k}_c^2 e^{-\tilde{k}_c^2} \frac{\sqrt{\tilde{k}_c^4+4 \tilde{v}_B^2 \tilde{k}_c^2}}{\tilde{k}_c^2+2 \tilde{v}_B^2} \int_{-1}^{1} \diff{\kappa}  \frac{ \tilde{B}(\tilde{k}_c \kappa)}{\sqrt{1-\kappa^2}},
\end{equation}
where 
\begin{multline}
    \tilde{B}(\tilde{k}_c \kappa)=J_{\nu_0}(\tilde{k}_c \kappa \tilde{a}_0) J_{\nu_0+\sum\limits_{\xi=1}^N (1+2\xi) (\nu_\xi-\nu'_\xi)}(\tilde{k}_c \kappa \tilde{a}_0) \\ \times\prod_{m=1}^N (-1)^{ m(\nu_m-\nu'_m)} J_{\nu_m}(\tilde{k}_c \kappa \tilde{a}_m)J_{\nu'_m}(\tilde{k}_c \kappa \tilde{a}_m)
\end{multline}
The integral cannot be solved analytically in general. Numerically we evaluate the expression for $N=5$. For our stirring setup the Fourier parameters are $\tilde{a}_\lambda=\frac{8\tilde{d}}{\pi^2(1+2\lambda)^2}$ and $\tilde{\omega}_\lambda=\frac{m \sigma^2}{\hbar}(2\lambda+1)\frac{\pi v}{2 d}=(2\lambda+1)\frac{\pi \tilde{v} \tilde{v}_B}{2 \tilde{d}}$, where we rescaled $\tilde{d}=\frac{d}{\sigma}$ and $\tilde{v}=\frac{v}{v_B}$.

%%%-Bibliography----%%%%
\bibliography{References}

\end{document}